\documentclass[conference]{IEEEtran}
\pdfoutput=1

\usepackage{amsmath, amssymb, subcaption, color, url, multirow, setspace, algorithm,  listings, balance, enumerate, graphicx, enumitem}
\usepackage{textcomp}

\usepackage{epstopdf}
\usepackage{booktabs}
\usepackage{algorithm}  
\usepackage{algpseudocode}  
\newcommand{\name}{{\tt{AuF}}}

\floatname{algorithm}{Process}

\usepackage{scalefnt}
 
\usepackage{tikz}

\def\BibTeX{{\rm B\kern-.05em{\sc i\kern-.025em b}\kern-.08em
    T\kern-.1667em\lower.7ex\hbox{E}\kern-.125emX}}

\begin{document}
\title{
Autonomous WiFi Fingerprinting for Indoor Localization
}

\author{Shilong Dai, Liang He, Xuebo Zhang}

\maketitle

\begin{abstract}
WiFi-based indoor localization has received extensive attentions from both academia and industry. However, the overhead of constructing and maintaining the WiFi fingerprint map remains a bottleneck for the wide-deployment of WiFi-based indoor localization systems. 
Recently, robots are adopted as the professional surveyor to fingerprint the environment autonomously. But the time and energy cost still limit the coverage of the robot surveyor, thus reduce its scalability.

%
%

To fill this need, we design an \underline{Au}tonomous WiFi \underline{F}ingerprinting system, called \name, which autonomously constructs the fingerprint database with time and energy efficiency. 
%
\name\ first conduct an automatic initialization process in the target indoor environment, then constructs the WiFi fingerprint database of in two steps: (i) surveying the site without sojourn, (ii) recovering unreliable signals in the database with two methods.
We have implemented and evaluated \name\ using a Pioneer 3-DX robot, on two sites of our $70$$\times$$90$m$^2$ Department building with different structures and deployments of access points (APs). The results show \name\ finishes the fingerprint database construction in 43/51 minutes, and consumes 60/82 Wh on the two floors respectively, which is a 64\%/71\% and 61\%/64\% reduction when compared to traditional site survey methods, without degrading the localization accuracy.
\end{abstract}





\section{Introduction}
\label{sec:introduction}
%
%
WiFi fingerprint-based localization systems --- using the signal strengths of WiFi APs to fingerprint the location from which the signal is collected --- have become the mainstream solutions for indoor localization.
These WiFi fingerprint-based localization methods consist of,  in general, two phases: an {\em offline} fingerprinting phase to construct the building's WiFi fingerprint map via site survey, and an {\em online} localization phase to position the mobile devices/users by checking the received WiFi signals with the fingerprint map~\cite{832252}. 

Significant research has been devoted to the online localization phase, achieving decimeter-level localization accuracy using advanced algorithms to match the online collected WiFi signals with the fingerprint map~\cite{Rajagopal:2018:EIS:3207947.3208003,tsui2009unsupervised,sun2014wifi,xie2016improved}.
However, a critical bottleneck of fingerprint-based indoor localization, i.e., the intensive overhead in constructing/maintaining the fingerprint map, remains unsolved: (i) an agent (e.g., a people carrying a WiFi scanner) needs to survey the building to collect data and construct the fingerprint map, and (ii) after construction, a fingerprint map needs to be updated frequently to mitigate the dynamics of WiFi signals~\cite{7174948}.

A variety of designs are proposed to use SLAM-enabled robots~\footnote{SLAM (Simultaneous Localization And Mapping)-enabled robot is able to construct/update a map of an unknown environment (e.g., a building) while simultaneously keeping track of the robot's location therein.} to facilitate the construction/maintenance of WiFi fingerprint map~\cite{lingemann2005high,biswas2012depth,varveropoulos2005robot}.
%
%
Most of these solutions use their robots in a ``travel-with-sojourn'' way: the robots visit, and stop at, each reference locations of the building to collect sufficient WiFi scans thereof, thus being able to fingerprint the reference locations reliably. 
The frequent stop of the robot, however, prolongs the time to finish the site survey (and hence fingerprint map construction) and moreover, increases the power consumption of the robot --- which are usually powered by batteries --- due to frequent de/acceleration, limiting the range the robots can cover/survey and impeding their deployments in large buildings.

To mitigate this deficiency, we design and implement an autonomous WiFi fingerprinting system, called \name, in which the robot constructs the WiFi fingerprint database by surveying the indoor environment without sojourn, thus expanding the range the robot can cover and shortening the time needed for fingerprinting.
\name's travel-without-sojourn, however, reduces the WiFi scans collected at specific reference locations, and thus degrades the reliability of collected WiFi measurements in the form of both lost and abnormal signals~\cite{chow2018efficient, bose2007practical}.
\name\ mitigates this degraded signal quality by using two novel signal recovery methods.
\begin{itemize}
    \item {\bf Lost Signal Recovery.~} \name\ recovers the lost signal using the strong correlation between 2.4GHz and 5GHz signals: (i) most commodity WiFi APs support both 2.4GHz/5GHz networking; (ii) for a given AP, the strength of the 2.4GHz/5GHz signal at a given location are strongly correlated; (iii) our empirical results show the two signal seldom lose at the same time. \name\ exploits this correlation between 2.4/5GHz signal to recover the lost signal during its site surveying, if anyone of them (but not both) is lost. 
    \item {\bf Abnormal Signal Recovery.~} \name\ detects abnormal WiFi samples based on a spatial model of signal strength. \name\ then uses the correlation between the current WiFi samples and the previously constructed database to recover them. 
    The key of this solution is that \name{}'s fingerprinting allows a short interval between fingerprint updates, making former information remain effective for the current site survey.  
\end{itemize}
Also note that \name\ recognizes the indoor environment and plans its site survey without requiring human operation, and thus being a fully autonomous system to construct/maintain the WiFi fingerprint database. 
The fingerprint database constructed by \name\ can then be used to build/update fingerprint maps by existing WiFi-based indoor localization systems.



We have evaluated \name\ on two floors of our Department building, as shown in Fig.~\ref{fig:robot}(b)(c). The results show \name\ to construct the fingerprint map with 64\%/71\% less time and 61\%/64\% less power on the two floors without degrading localization accuracy, when compared to the traditional site survey method~\cite{mirowski2012depth,nguyen2016low}.

\begin{figure}
\centering
\includegraphics[width=1\linewidth]{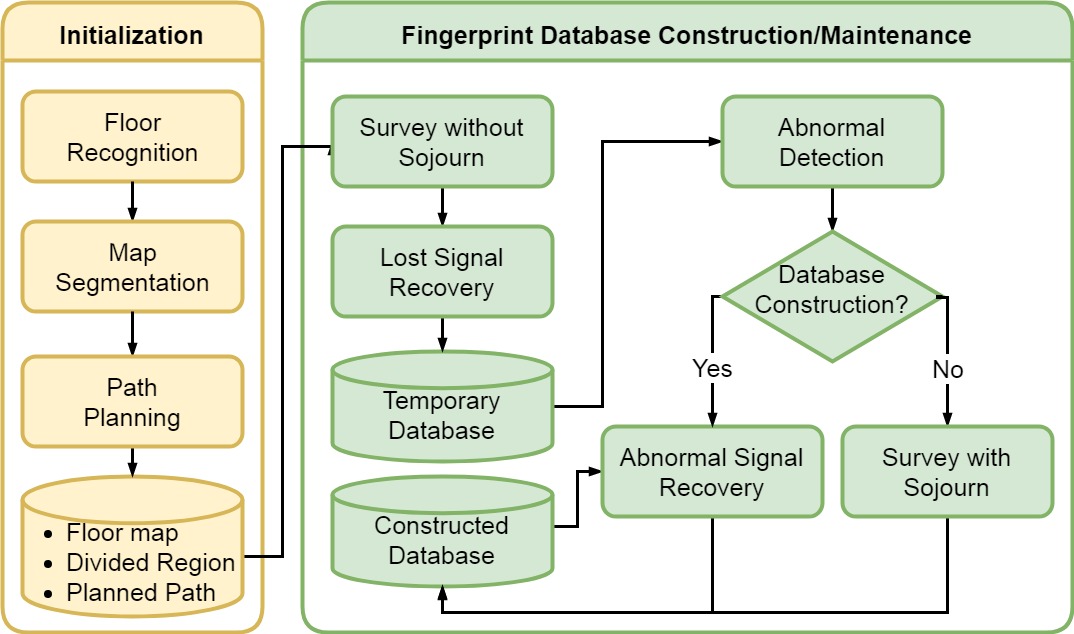}
\caption{Overview of \name: constructing the WiFi fingerprint map via autonomous floor recognition, motion planning, and then fingerprint map construction with signal recovery.}
\label{fig:overview}
\end{figure}

\section{Overview}
\label{sec:overview}

%

Fig.~\ref{fig:overview} depicts an overview of \name. First, \name\ conducts an automatic initialization process. It acquires a building's floor map for the robot's localization and navigation. Then the floor map is segmented to smaller regions with regular shapes, which is essential for an efficient path planning. After map segmentation, the traveling paths of the robot is planned for the site survey. 

Survey-without-sojourn is the beginning of \name's fingerprint database construction, during which the robot surveys the floor without sojourn to build a temporary fingerprint database in a short time. Then, lost signal recovery is performed by exploiting the correlation between 2.4/5GHz signals. To further improve the reliability of this signal recovery, \name\ also refines the temporary database by recovering abnormal signals.  

These abnormal signals are identified by \name's abnormal detection module. Then, we determine whether these signals can be recovered from the previous fingerprints. For fingerprint database construction, i.e. no past information, the robot surveys with sojourn at locations of abnormal signals, and acquire multiple samples for every location to complete the fingerprints for the temporary fingerprint database. 

%
Otherwise, for fingerprint database maintenance, we recover abnormal signals by exploiting the pattern of signals’s short-term correlation. After this database refinement, the temporary fingerprint database is concluded as the constructed fingerprint database.

\section{Initialization}
\label{sec:floor_recognition}

\name{} needs to do some preparations before its site survey, which  needs to be conducted only once. Also, this process is automatic thanks to the robotics technique, thus making \name{} easy to deploy.
The first task of \name's initialization is to discover the floor map of the floor-of-interest, which is the prerequisite for the robot's navigation indoors. Next, \name's divides the floor map into regions with regular shapes. Then in each region, \name\ plans the surveying paths for its robot.

\begin{figure*}[t]
\centering
\begin{subfigure}[t]{0.24\linewidth}
\includegraphics[width=1\linewidth]{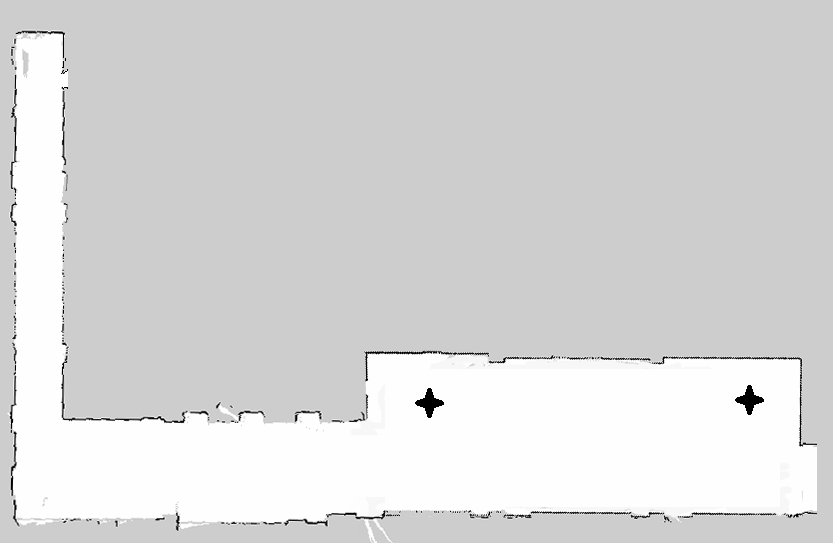}
\caption{The raw floor map identified using laser. The two marks are explained in Sec.\ref{sec:signal_recovery}}
\end{subfigure}
\hfill
\begin{subfigure}[t]{0.24\linewidth}
\includegraphics[width=1\linewidth]{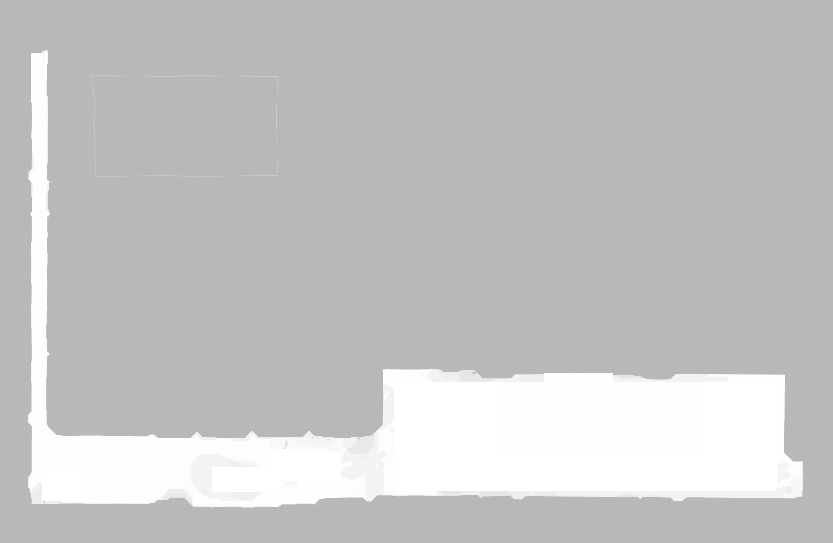}
\caption{The reduced floor map in which the robot can travel freely.}
\end{subfigure}
\hfill
\begin{subfigure}[t]{0.24\linewidth}
\centering
\includegraphics[width=1\linewidth]{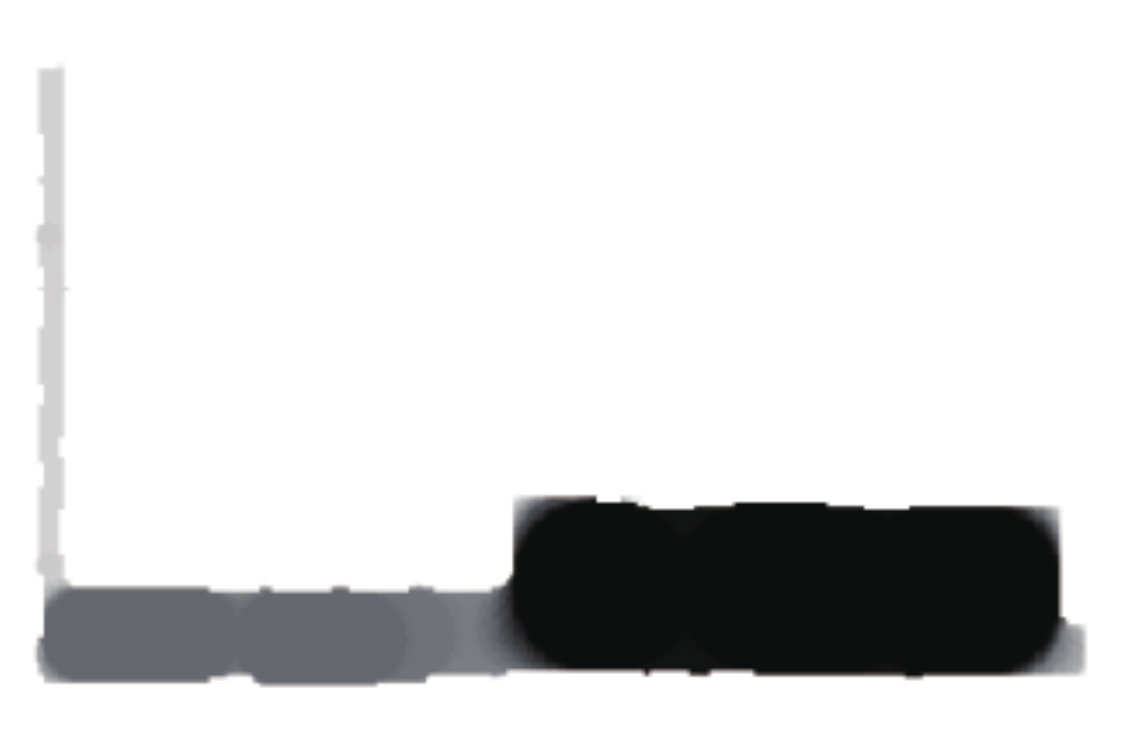}
\caption{The darker the pixel,
the highest its value.}
\end{subfigure}
\hfill
\begin{subfigure}[t]{0.24\linewidth}
\centering
\includegraphics[width=1\linewidth]{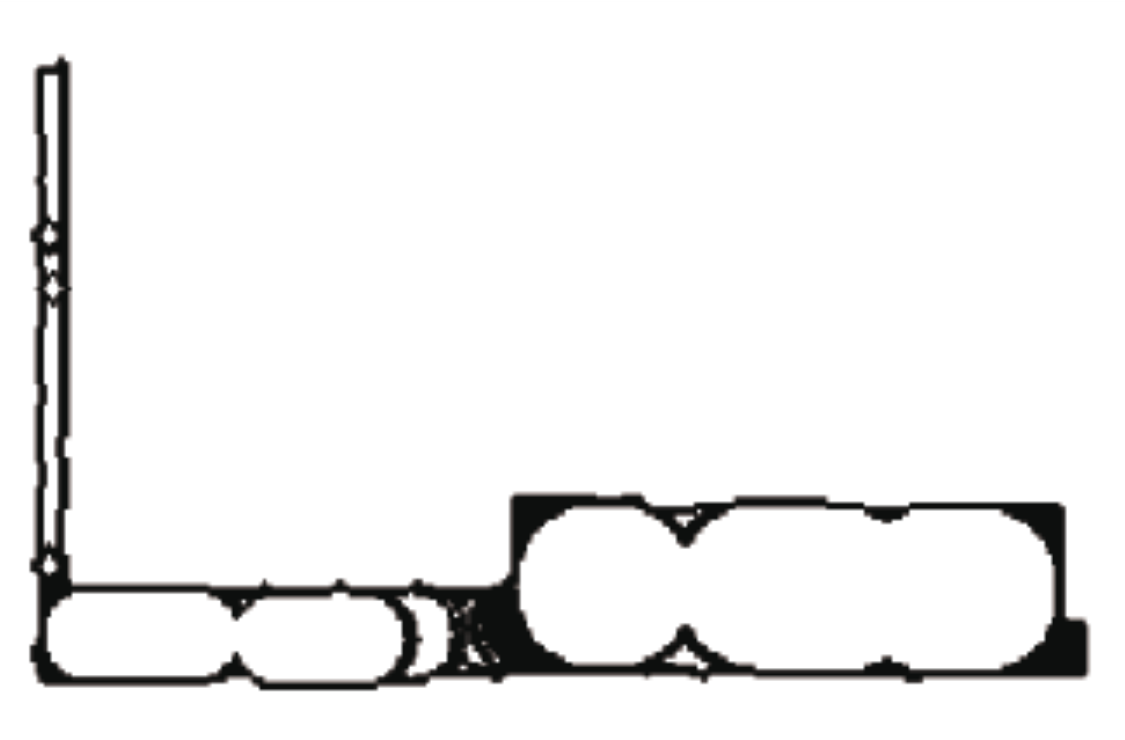}
\caption{Adjacent pixels with same value are aggregate into a region.}
\end{subfigure}
\\
\begin{subfigure}[t]{0.24\linewidth}
\centering
\includegraphics[width=1\linewidth]{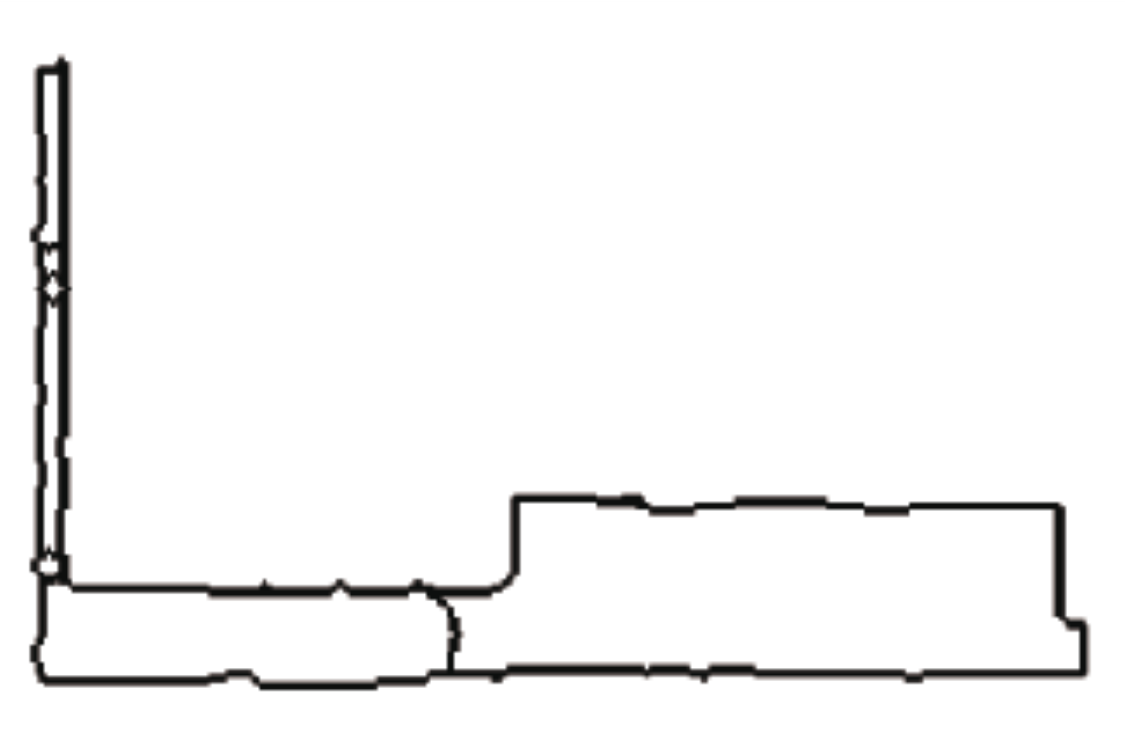}
\caption{Small ripples in Fig.~\ref{fig:path_planning}(d) are removed.}
\end{subfigure}
\hfill
\begin{subfigure}[t]{0.24\linewidth}
	\centering
	\includegraphics[width=1\linewidth]{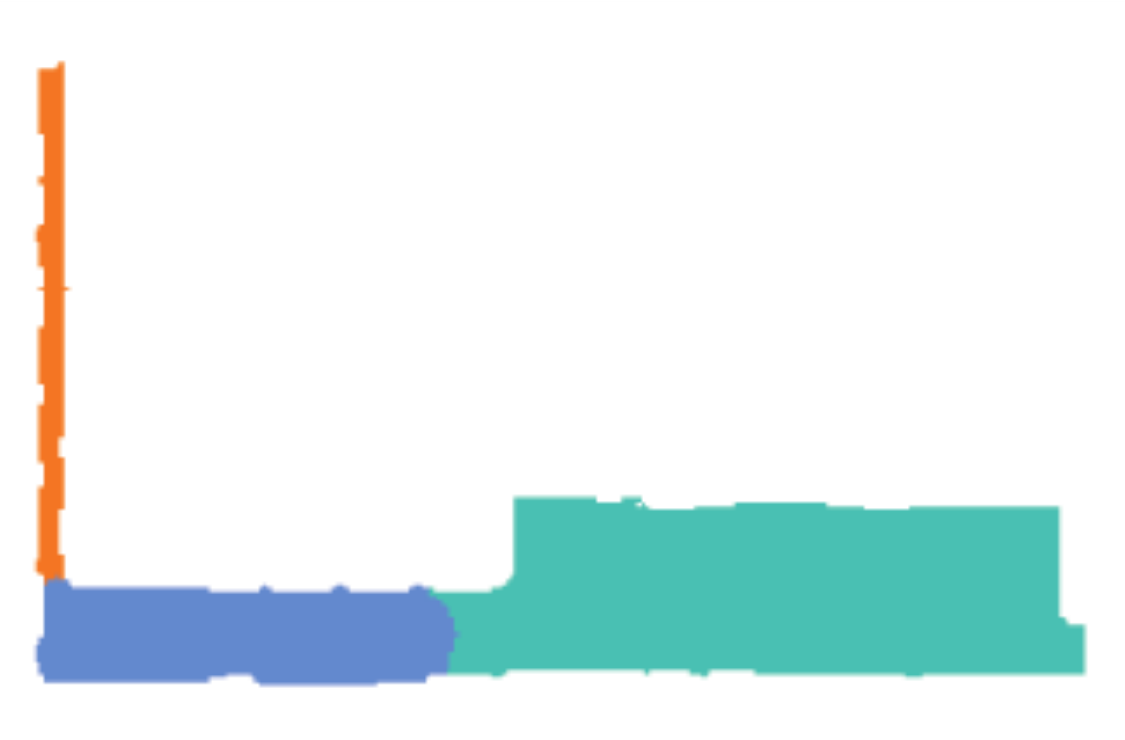}
	\caption{Regions with similar value in Fig.~\ref{fig:path_planning}(e) are merged, and final regions are founded.}
\end{subfigure}
\begin{subfigure}[t]{0.24\linewidth}
	\centering
	\includegraphics[width=1\linewidth]{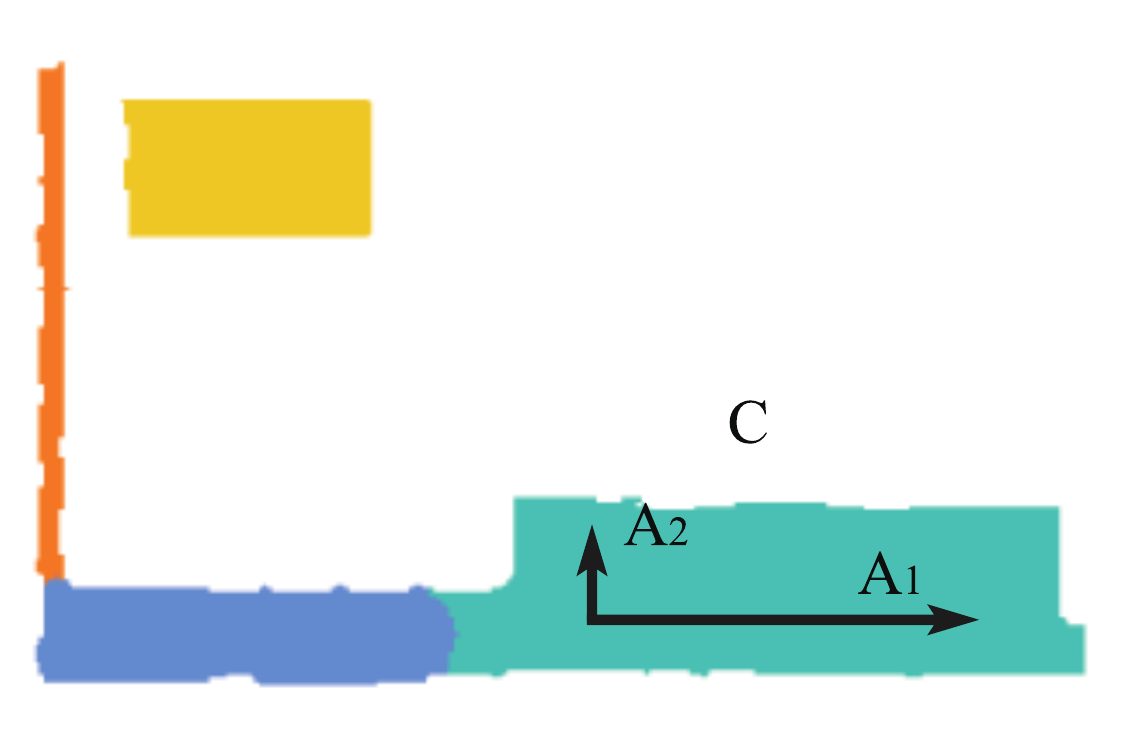}
	\caption{Each region's direction is determined by its shape.}
\end{subfigure}
\hfill
\begin{subfigure}[t]{0.24\linewidth}
	\centering
	\includegraphics[width=1\linewidth]{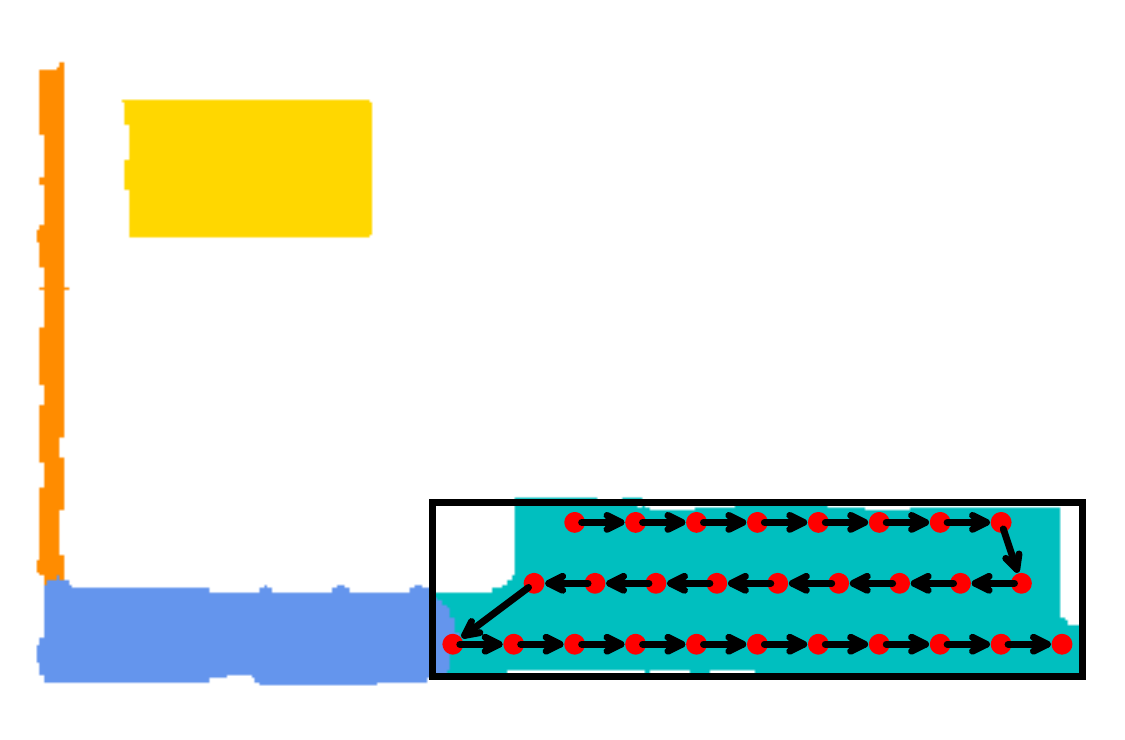}
	\caption{The robot needs to visit red dots according to regions' principal axis.}
\end{subfigure}
\caption{The whole process of initialization. }
\label{fig:path_planning}
\end{figure*}

\vspace{+3pt}
\noindent $\bullet$ {\bf Floor Map Discovery.~}
\name\ discovers the floor map using a SLAM-enabled robot, as shown in Fig.~\ref{fig:robot}. The robot scans its surrounding environment with laser while surveying the building, and constructs the building's floor map based on the scanning results. The spanning-tree algorithm~\cite{gabriely2001spanning} is adopted to achieve an automatic SLAM process.
A gray-scale grid map (e.g., as shown in Fig.~\ref{fig:path_planning}(a)) is obtained after the site survey, in which the pixels with gray-scales smaller than a pre-defined threshold represents the obstacles of the building (e.g., walls). 
\name\ then increases the obstacle area by the size of the robot, thus obtaining the floor area where the robot can travel freely, as shown in Fig.~\ref{fig:path_planning}(b).

%
%


%

%

\vspace{+3pt}
\noindent $\bullet$ {\bf Map Segmentation.~}
Instead of directly planning paths for the entire floor, \name{} first divides the above-recognized floor into subareas before planning paths for the robot, because of two reasons. First, directly planning path covering a irregular space is a challenging, while identifying the optimal survey of a regular space is much more feasible --- the map segmentation facilitates planning paths to efficiently cover a irregular space \footnote{By ``efficient", we mean the paths have the least turning/overlapping}. 
%
%
The second reason necessities \name's map segmentation is the computational cost of constructing the fingerprint database, which grows in cubic with the number of collected samples. In fact, site surveys are  conducted separately in each region to limit the size of samples during once site survey. The computation complexity of \name{} and its solution will be discussed in detail in Sec.~\ref{sec:discussion}.

\name\ grounds its map segmentation using MAORIS~\cite{mielle2018method}. Below we briefly describe its major steps:
First, a free space image, whose pixel value represents the size of the region it belongs to, is created based on the distance image of the floor map\footnote{Pixels in the distance image represent Euclidean distance to the nearest obstacle}. The free space image is initiated with an empty image. A circular mask is created for each pixel of the distance image and centered on the pixel, whose radius are determined by the value of the pixel. Then for pixels in the circular mask, if the value of the equivalent pixel in the free space image is less than the circle radius, the value of the pixel in the free space image is changed to the radius. The result is showed in Fig.~\ref{fig:path_planning}(c).
Then, in the free space image, adjacent pixels with same value are grouped into regions, which can be seen in Fig.~\ref{fig:path_planning}(d).
Clearly, Fig.~\ref{fig:path_planning}(d) is over-segmented, and some regions are too small to represent a semantic area. MAORIS removes these ripples with a simple rule: a region is incorporated into its neighbor if they overlap for more than 40\%, as visualized in Fig.~\ref{fig:path_planning}(e).
After removing ripples, some regions are still pieces of the same place like the left corridor in Fig.~\ref{fig:path_planning}(e), resulted from slightly different distances to the obstacle. These regions with similar value are merged, and Fig.~\ref{fig:path_planning}(f) plots the final segmented map.

\vspace{+3pt}
\noindent $\bullet$ {\bf Path Planning.~}
After segmenting the floor map, \name\ needs to determine the principal axis direction for each region in the map, moving along which requires the minimum number of robot's turning. 
Let us consider Fig.~\ref{fig:path_planning}(g) as an example. Denoting the coordinates of the pixels in one region ${\rm{C}}$ as ${P}$, we get a set of coordinates representing that area:
\begin{equation}
{\rm{C}} = \left\{ {P_1}, {P_2}, \cdots, {P_l} \right\}.
\end{equation}
We calculate the variance of all pixels in $C$ by:
\begin{equation}
Var(C)
=\begin{bmatrix}
cov(x,x) &cov(x,y)\\ 
cov(y,x) & cov(y,y)
\end{bmatrix},
\end{equation}
where $x, y$ denotes the pixels' coordinates.
Because $Var(C)$ is a real symmetric matrix, its two eigenvectors (denoted as $\rm A_1$ and $\rm A_2$, as shown in Fig.~\ref{fig:path_planning}) are orthogonal, representing the two directions of the spatial distribution of the region's elements. So, the $\rm A_1$ with a higher eigenvalue is the principal axis of region $\rm{C}$.


Next, \name\ discretizes each region into grids (we use $0.8m\times0.8m$ grids) according to the corresponding principal axis direction, and combines the grid centers as the traveling paths of the robot.
To acheive this, \name\ identifies the minimum rectangles covering each of the region. Denoting $F$ as the world coordinate system, ${{}^F{P}}$ as the pixel point in $F$. 
\name\ uses the two directions $\rm A_1$ and $\rm A_2$ as the region's coordinate system (i.e., {$F'$} in Fig.~\ref{fig:path_planning}).
The rotation transformation from ${{}^{F'}{P}}$ to ${{}^F{P}}$ is thus:
\begin{equation}
{^{F'}{P}}=\left[ {\begin{array}{*{20}{c}}
{\rm {A_1}}^\mathrm{T}\\
{\rm {A_2}}^\mathrm{T}
\end{array}} \right]\cdot {{}^F{P}}.
\end{equation}
\name\ then finds region-C's max/minimum coordinates according to {$F'$}, thus identifying a minimum rectangle covering the area of region-C.
At last, \name\ divides each of the above-identified rectangles into grids according to the coordinate system {$F'$}, and ask the robot to pass the grids' center in boustrophedon like Fig.~\ref{fig:path_planning}(h). 
%

\begin{figure*}[t]
    \centering
    \begin{subfigure}[t] {0.25\linewidth}
    \centering
    \includegraphics[width = 1\linewidth]{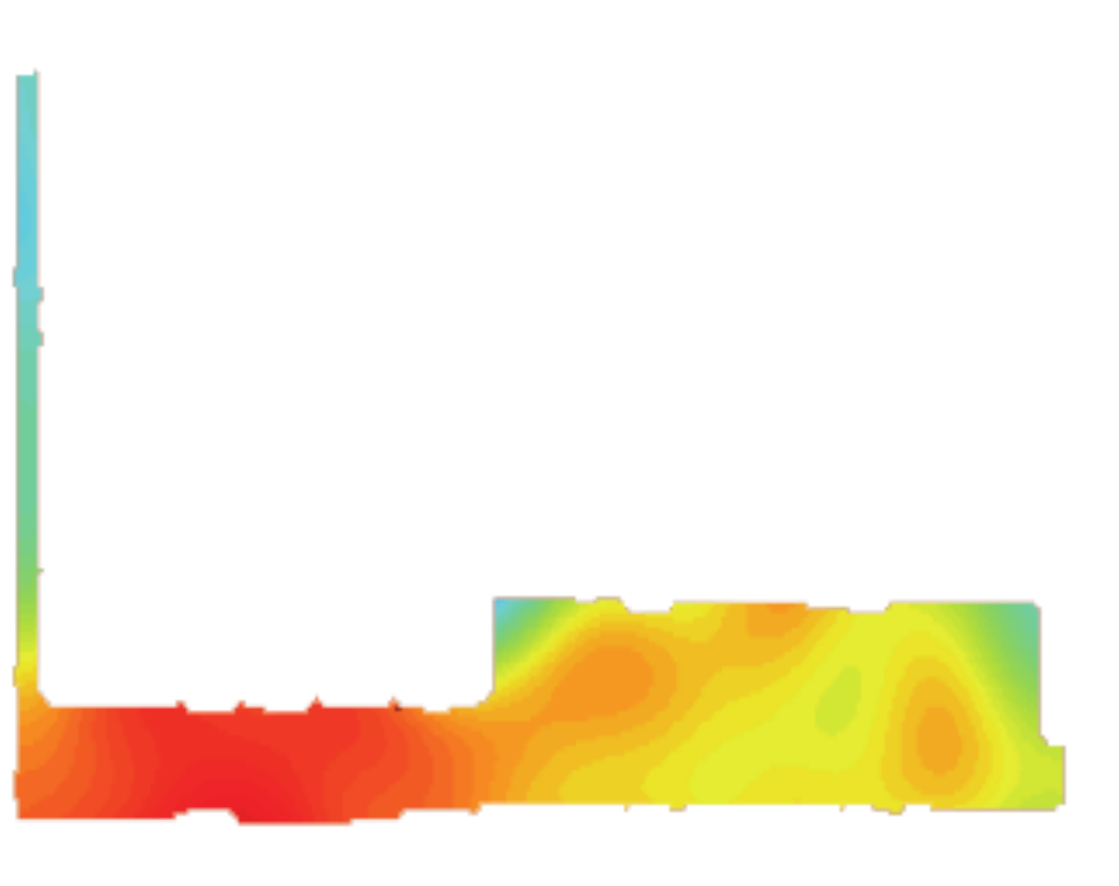}
    \caption{Signals collected during 9s sojourn.}
    \end{subfigure}
    \hfill
    \begin{subfigure}[t] {0.25\linewidth}
    \centering
    \includegraphics[width = 1\linewidth]{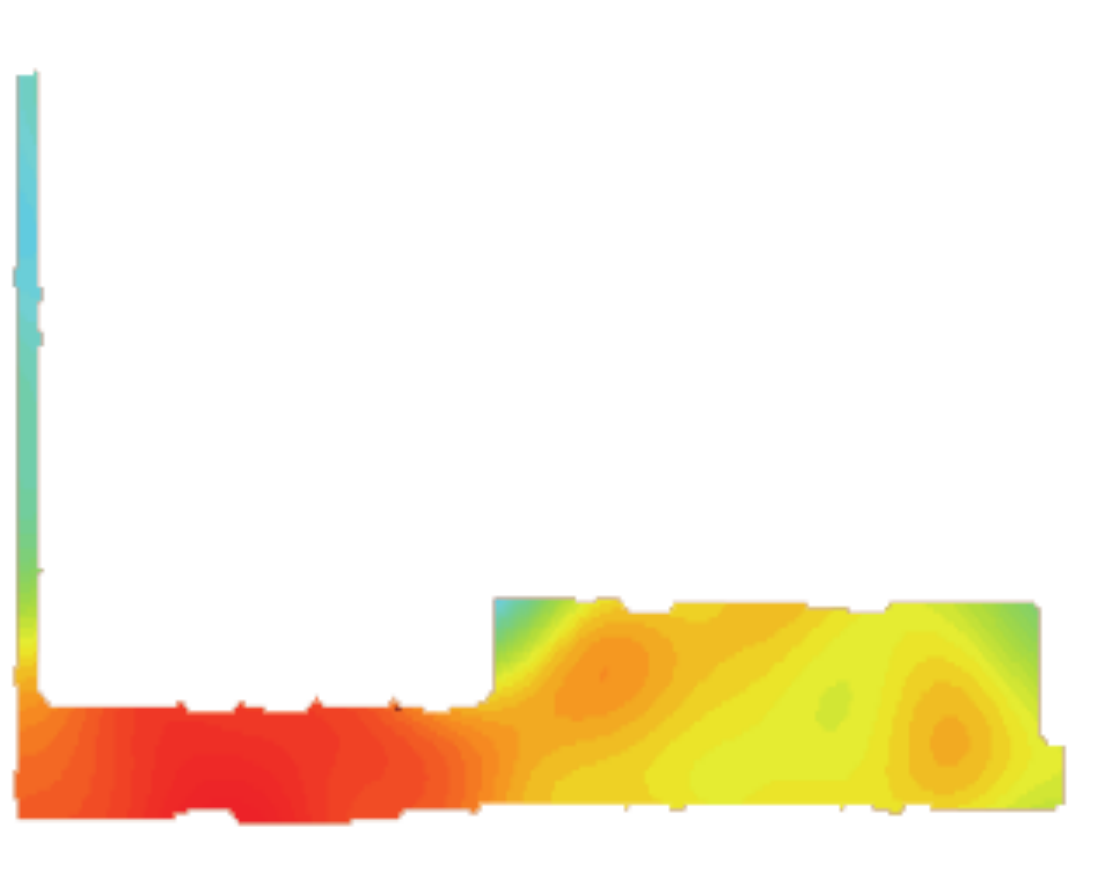}
    \caption{Signals collected during 3s sojourn.}
    \end{subfigure}
    \hfill
    \begin{subfigure}[t] {0.25\linewidth}
    \centering
    \includegraphics[width = 1\linewidth]{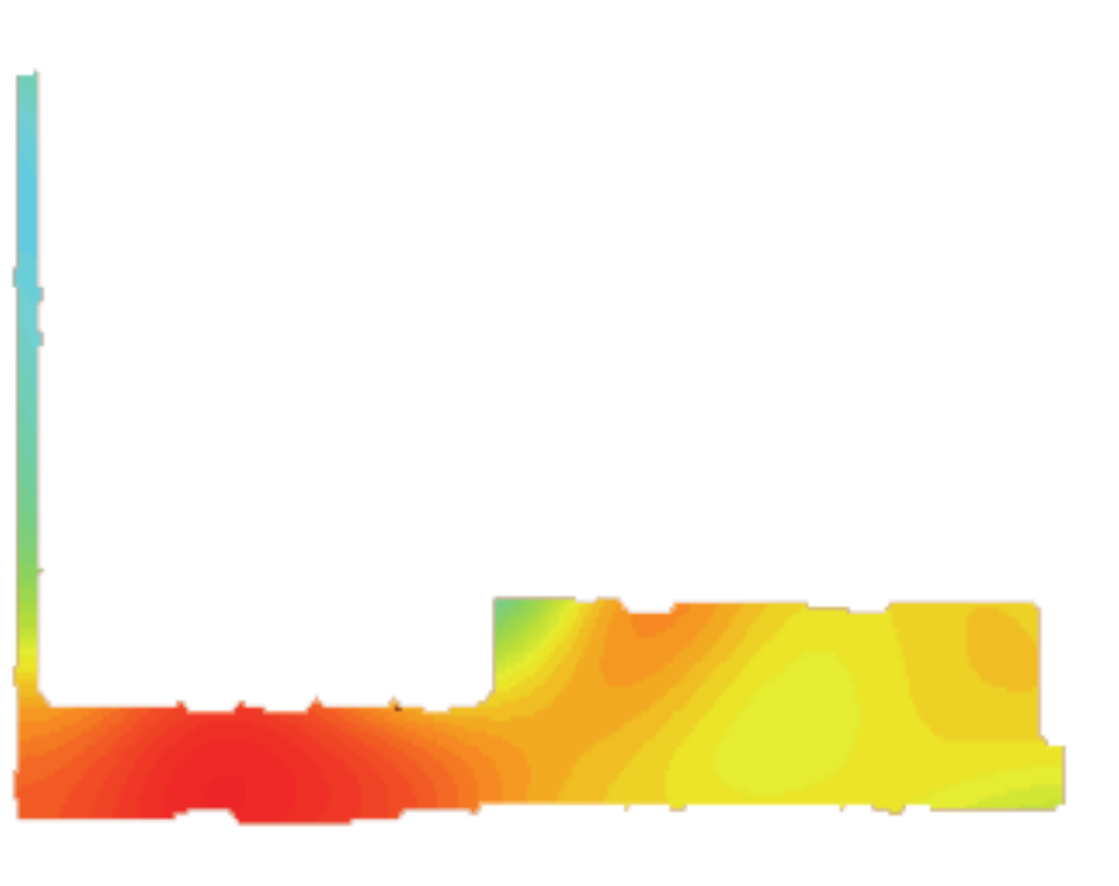}
    \caption{Signals collected during the movement.}
    \end{subfigure}
    \hfill
    \begin{subfigure}[t] {0.2\linewidth}
    \centering
    \includegraphics[width = 1\linewidth]{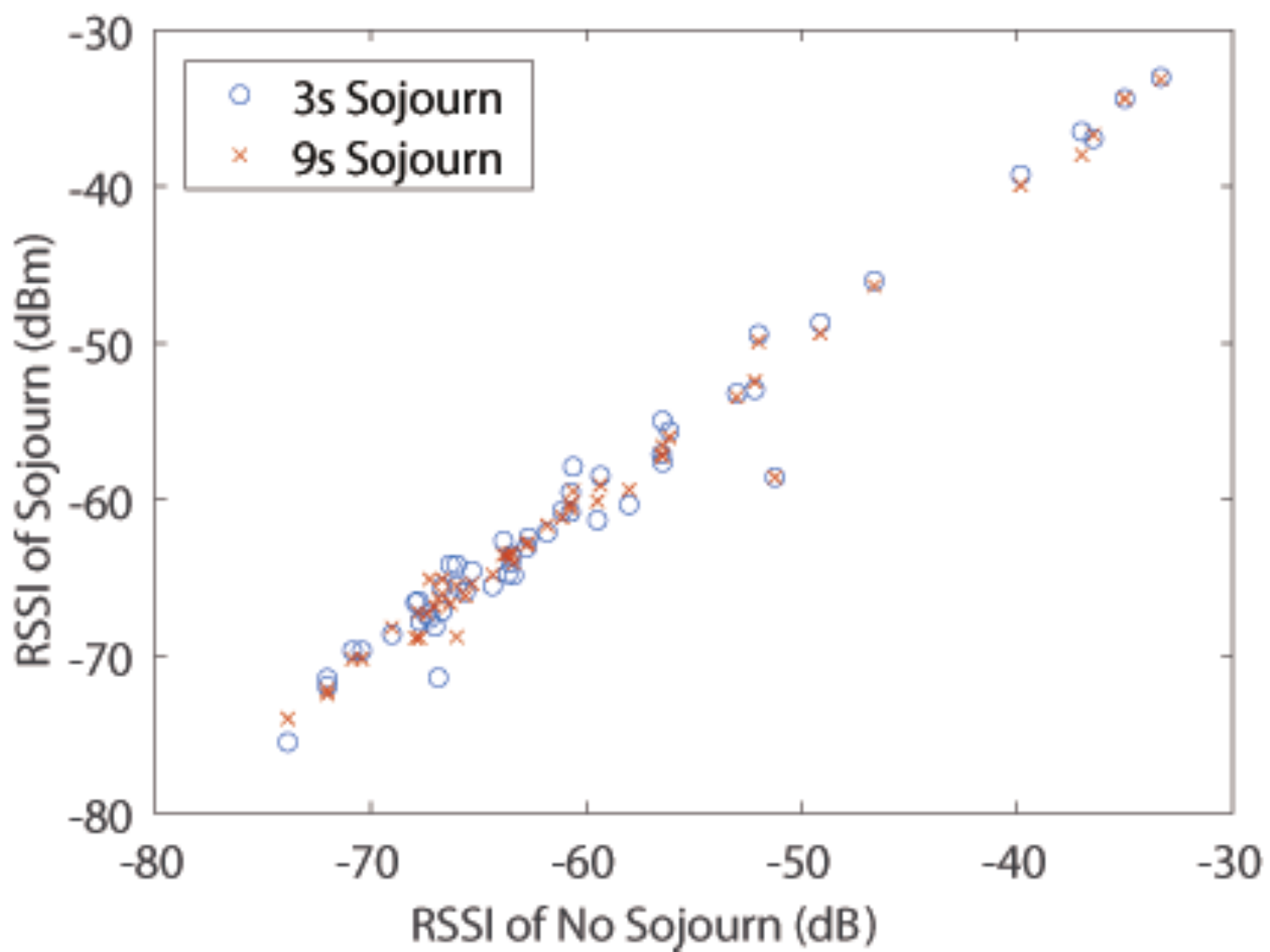}
    \caption{Relationship between signals collected with or without sojourn.}
    \end{subfigure}
    \caption{No clear dependency between robot's travel speed and the received WiFi signals is observed.}
    \label{fig:speed}
\end{figure*}

\section{Fingerprint Database Construction}
\label{sec:survey_process}

In this section we describe \name's construction of the fingerprint database. First we verify the feasibility of continues movement of the robot during collecting fingerprints. Then we explain the dual-band (i.e., 2.4GHz and 5GHz) signal recovery and the dependent signal model thereof.

\subsection{Survey without Sojourn}
\label{sec:fast_survey}


Unlike traditional systems~\cite{ocana2005indoor,youssef2005horus,832252} requesting the surveyor/robot stopping at locations and scanning WiFi multiple times, the robot in \name\ travels through generated grids without sojourn to shorten the survey process.
%

%
%
\name's survey-without-sojourn has two challenges.
Intuitively, the robot's continuous movement may affect the collected WiFi signals due to Doppler shift.
%
To closely examine this, we use the robot to survey the area in Fig.~\ref{fig:path_planning}(a) to collects the WiFi signal from a given AP.
We collect three datasets, data collected with a 9s sojourn at each grid, with a 3s sojourn at each grid, and without sojourn.
Fig.~\ref{fig:speed}(a)(b)(c) visualize the heatmaps of the three dataset, and Fig.~\ref{fig:speed}(d) uses a scatter plot to represent relationships of signals collected with or without sojourn, showing no clear dependency of signal strength with the movement.
This is likely due to the relatively slow travel speed of the robot, e.g., 0.5m/s in the above measurements.
Another challenge is even more severe: with one fingerprint per location, how to ensure that this fingerprint can reliably represent the local signal strength. \name\ identifies, and then improves, the locations whose signal representation is unreliable. The following secs.~\ref{sec:signal_recovery} and \ref{sec:signal_supplement} separately explains separately two kind of unreliable signals and the corresponding remedies.

\subsection{Recovery of Lost Signal}
\label{sec:signal_recovery}

\begin{figure}[htbp]
    \centering
    \begin{minipage}{1\columnwidth}
    \begin{subfigure}{.49\columnwidth}
    \includegraphics[width=1\linewidth]{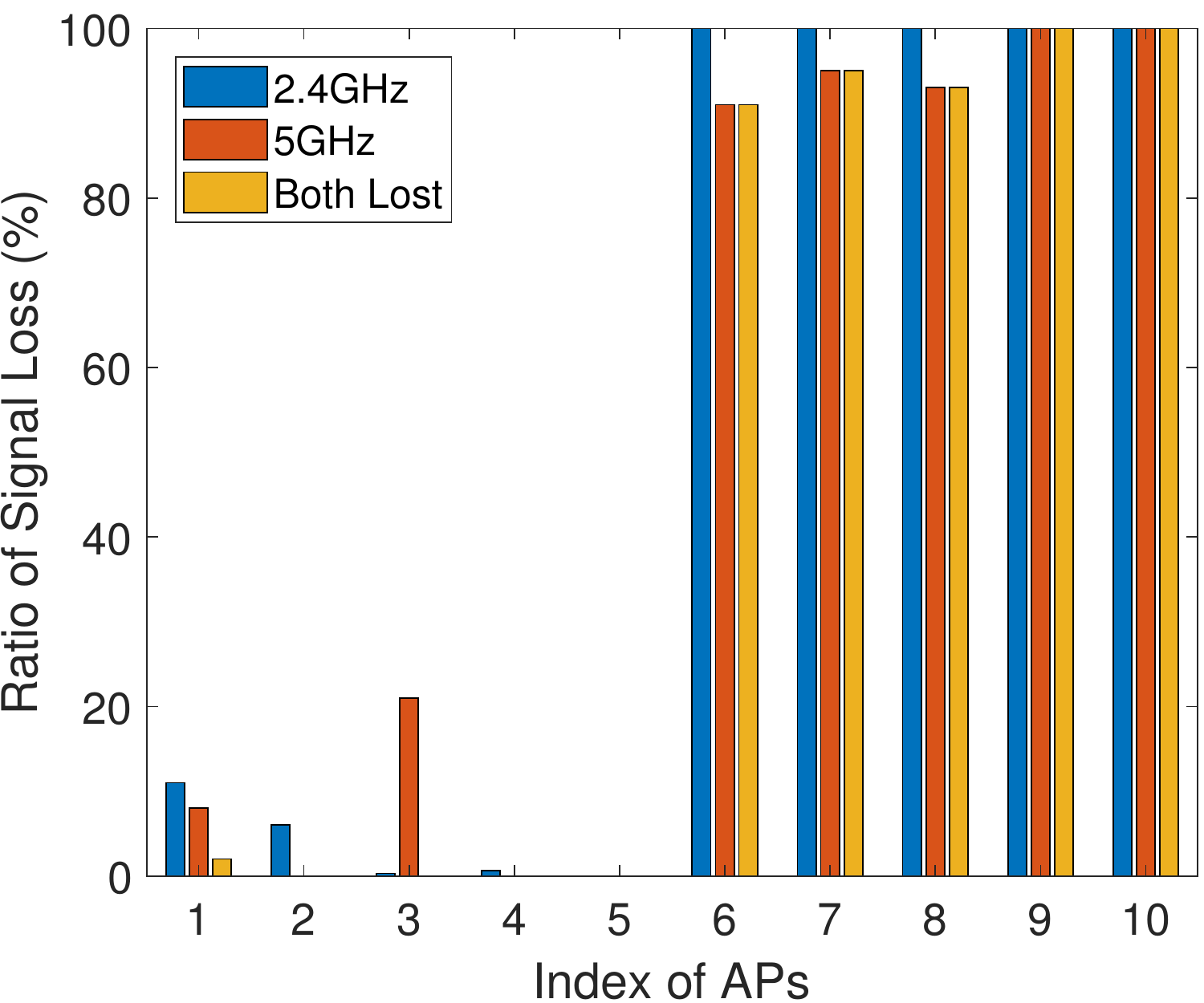}
    \caption{}
    \end{subfigure}
    \hfill
    \begin{subfigure}{.49\columnwidth}
    \includegraphics[width=1\linewidth]{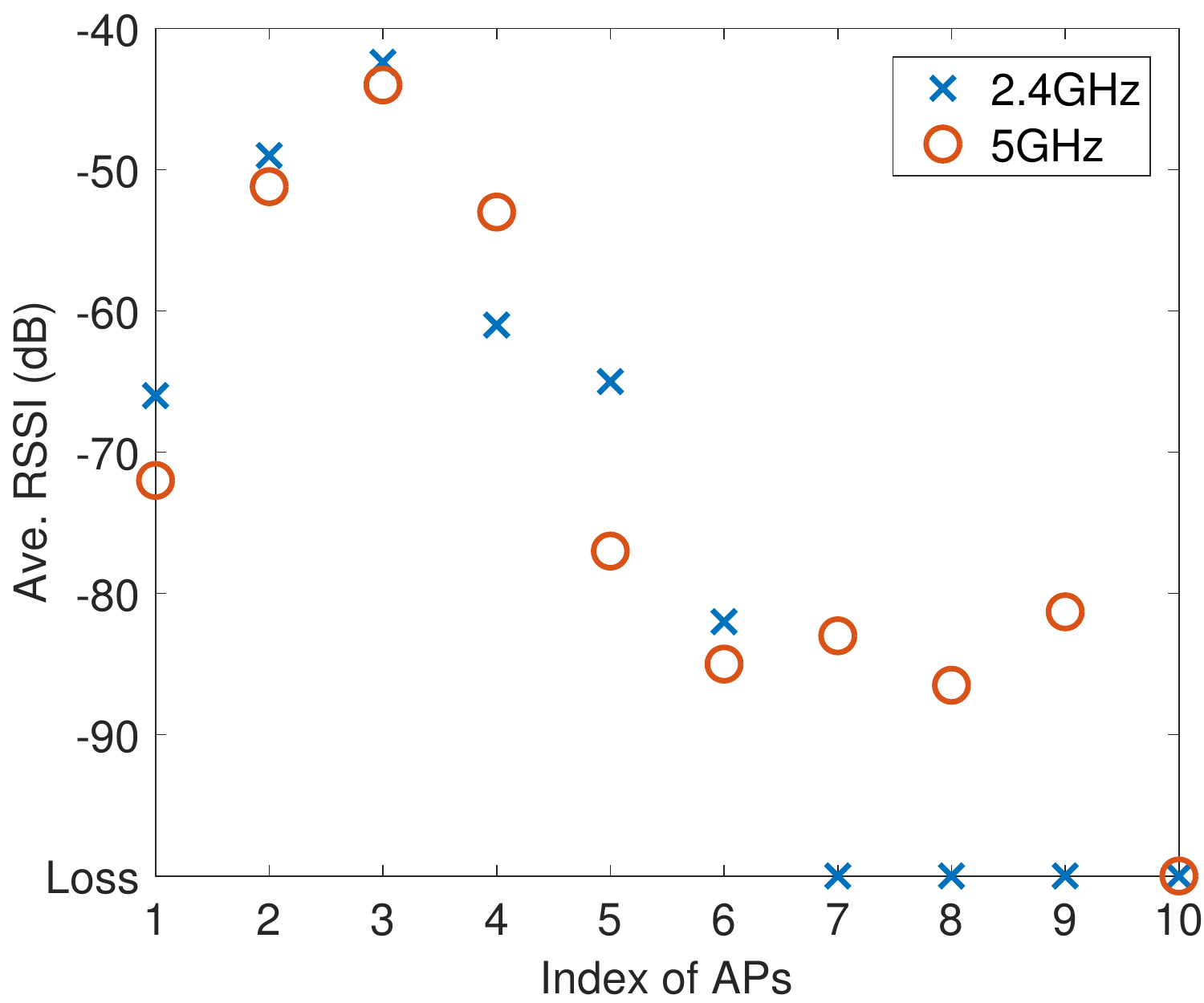}
    \caption{}
    \end{subfigure}
    \caption{Loss ratio and the signal strength of WiFi signals received at a fixed location.}
    \label{fig:loss_24and5}
    \end{minipage}
\end{figure}


\vspace{+3pt}
\noindent $\bullet$ {\bf Signal Loss.~}
%
The first type of unreliable signals is random signal loss. Lost signals are common during the survey process, which are usually indicators of long distances to the AP and poor signal strengths. We set the RSSIs of missed signals a low value (i.e. -100dBm).
%
%
Random signal loss occurs even when APs are just nearby, caused due to a variety of reasons such as obstruction of APs, scanning duration.

\begin{figure}[htbp] 
		\centering
        \begin{subfigure}[t]{1\linewidth}
        \centering
        \includegraphics[width=1\linewidth]{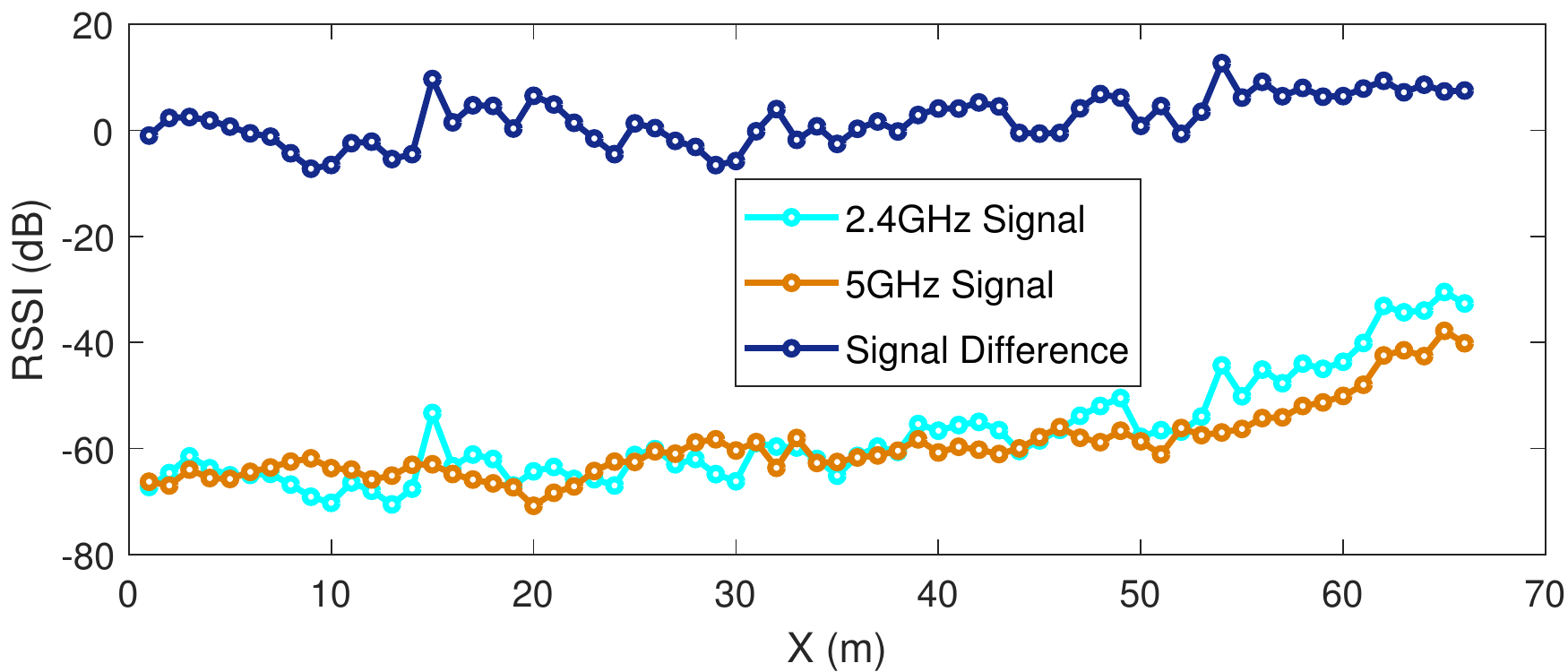}
        \end{subfigure}
		\caption{The two signals received at the same location show clear correlation.}
		\label{fig:24and5GHz}  
\end{figure}

Unlike traditional survey method, \name{} does not rely on multiple collected samples at a location to mitigate the random signal loss. Instead, \name\ recovers the lost signal by exploiting the correlation between 2.4GHz signals and 5GHz signals from the same physical AP.
To corroborate feasibility of this signal recovery, we collect the dual-band WiFi signals from the 10 APs on the 6th floor at two fixed locations, marked in Fig.~\ref{fig:path_planning}, for about 2 hours.
%
Fig.~\ref{fig:loss_24and5} summarizes the ratios of signal losses, and Fig.~\ref{fig:loss_24and5} plots the signals' average RSSI. 
Comparison of Figs.~\ref{fig:loss_24and5} and \ref{fig:loss_24and5} shows: (i) loss is observed at both 2.4GHz and 5GHz signals; (ii) the loss of a single frequency signal from a given AP is not necessarily caused by too weak a signal (e.g., with AP4 and AP5); (iii) the loss of both 2.4GHz and 5GHz signals from a given AP, however, does indicate a weak signal strength (e.g., with AP7--AP10).
\vspace{+3pt}
\noindent $\bullet$ {\bf Signal Correlation.~}
\name\ recovers the lost signal using the correlation between 2.4GHz and 5GHz signals.
To corroborate such a signal correlation, we evenly select 66 locations along a 65m straight path, and stop the robot 10s at each location to collect the 2.4GHz and 5GHz signals of an AP located at the 67m location. The average RSSI at each point is plotted in Fig.~\ref{fig:24and5GHz}. 
The two traces of WiFi signals have a correlation coefficient of $0.92$, implying the feasibility to recover the lost 2.4GHz signal based on the 5GHz signal, and vice versa. Furthermore, we find that the spatial distribution of the difference between the two signals is regular, as shown in Fig.~\ref{fig:24and5GHz}.

The correlation between 2.4GHz and 5GHz signals can be explained analytically. 
According to the log-normal shadowing model~\cite{rappaport1996wireless}, the RSSI of wireless signals can be expressed as
\begin{eqnarray}
	P(d) &=& 10\cdot {\log _{10}}(\frac{{{W}{G_{AP}}{G_{MT}}\lambda ^2}}{{16{\pi ^2}d_0^2L}}) 
	- 10\cdot \beta \cdot  {\log _{10}}(\frac{d}{{{d_0}}}) \nonumber \\
    &&+ X(0,{\delta ^2})
    \label{eq:shadow_log}
\end{eqnarray}
where $P(d)$ is the RSSI measured at a distance $d$ from a given AP, with transmission power $W$.
$G_{AP}$ and $G_{MT}$ are the antenna gains on the AP and the mobile terminal, respectively. $L$ is the system's loss factor, $\lambda$ is the carrier's wavelength, $\beta$ is the path loss exponent, and $X(0,{\delta ^2})$ is a zero-mean Gaussian distributed random variable, capturing the shadowing effect~\cite{rappaport1996wireless}.

\begin{figure}[t]
    \centering
    \includegraphics[width=1\linewidth]{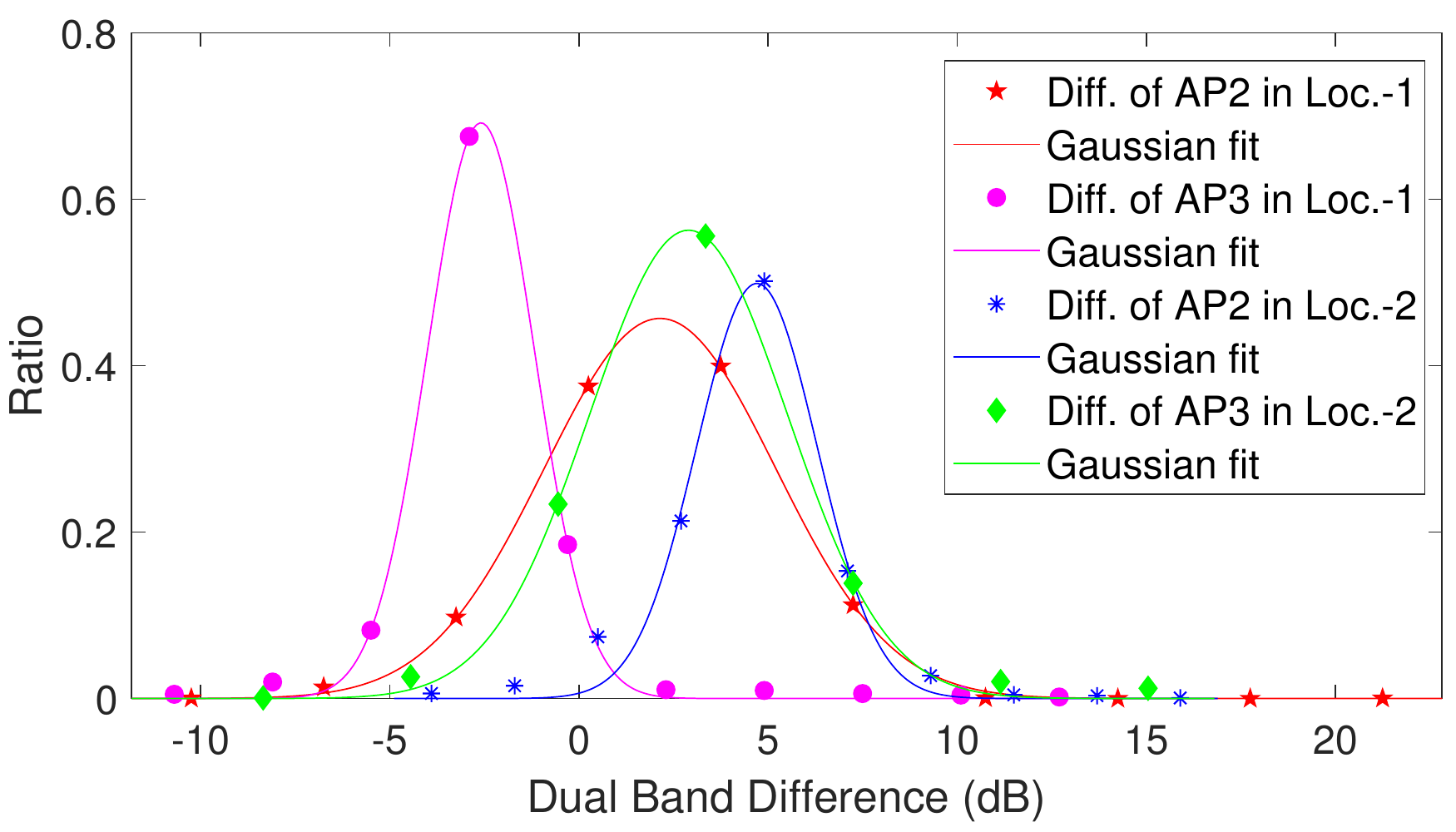}
    \caption{Fitting the difference between 2.4GHz and 5GHz signals as Gaussian.}
    \label{fig:fit}
\end{figure}

Denote ${P_{2.4}}(d)$ and ${P_{5}}(d)$ as the signal strength of the 2.4GHz and 5GHz signal measured at a distance $d$ of the given AP, respectively. The difference between the two signals' RSSI can be calculated based on Eq.~\ref{eq:shadow_log} as
{
\begin{equation}
\begin{aligned}
    &{P_{2.4}}(d) - {P_5}(d)= 10\cdot {\log _{10}}(\frac{{\lambda _{2.4}^2}}{{\lambda _5^2}}) 
    \\
    &- 10({\beta _{2.4}} - {\beta _5}){\log _{10}}(\frac{d}{{{d_0}}})
     + X(0,\delta _{2.4}^2) - X(0,\delta _5^2),
\end{aligned}
\label{eq:difference}
\end{equation}
}
%
which can be further simplified to
\begin{equation}
	{P_{2.4}}(d) - {P_5}(d)= f(d)+X(\mu ,{\delta ^2}),
	\label{eq:simplifieddifference}
\end{equation}
where 
\begin{equation}
f(d) = -10\cdot ({\beta _{2.4}} - {\beta _5})\cdot {\log _{10}}(\frac{d}{{{d_0}}}).
\end{equation}
Eq.~(\ref{eq:simplifieddifference}) implies that the RSSI difference of 2.4GHz and 5GHz signals at the same location can be approximated as a Gaussian variable, thus explaining their correlation. 
Fig.~\ref{fig:fit} plots the results when fitting the difference between the 2.4GHz and 5GHz signals in Fig.~\ref{fig:loss_24and5} as Gaussian, showing high fitting goodness and thus verifying the above reasoning. Note the signals received from AP2 and AP3 are used here because of their relatively low loss ratios (and thus sufficient samples).

\vspace{+3pt}
\noindent $\bullet$ {\bf Signal Recovery.~}
\name, inspired by the correlated signals, recovers the lost 2.4GHz signal based on the 5GHz signal collected at the same scan, and vise versa. 
%
%

\begin{figure}[htbp]
\centering
\includegraphics[width=0.98\linewidth]{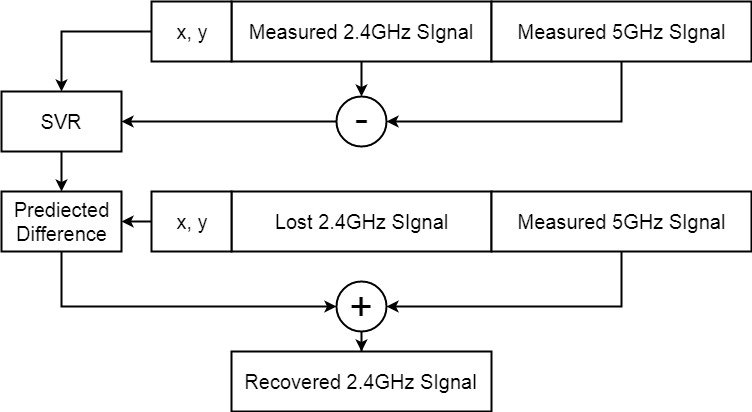}
\caption{Flow chart of \name's signal recovery, with recovering the lost 2.4GHz signals using received 5GHz signals as an example. }
\label{fig:recovery_model}
\end{figure}

We use the recovery of lost 2.4GHz signals with 5GHz signals to walk through \name's signal recovery. 
%
%
%
Eq.~(\ref{eq:simplifieddifference}) indicates the difference between the 2.4GHz and 5GHz signals consists of $f(d)$ and a Gaussian noise, where $f(d)$ is a function of the signal's propagation distance $d$. 
Inspired by this, \name\ recovers the lost 2.4GHz signal by training a SVR (Support Vector Regression) model for each AP, with the location's 2-D coordinates as input and the signal difference thereat as output.
%
%
Fig.~\ref{fig:recovery_model} summarizes such a signal recovery process of \name.
%


Clearly, \name's signal recovery requires at least a valid signal (i.e., either 2.4GHz or 5GHz) is received. 
In case of both signals are lost, \name\ will use a weak signal (e.g., assuming a -100dBm RSSI) to fingerprint that location, inspired by the empirical observation uncovered in Figs.~\ref{fig:loss_24and5}(a)(b).

\section{Recovery of Abnormal Signal}
\label{sec:signal_supplement}

Refinement of the fingerprint database is needed to mitigate abnormal signal caused due to system noise in the temporary database. In the building, the fluctuation of the wireless signal sometimes gets high due to the multi-path effect. Also occasional disturbances can increase the inaccurate signal strength in the database. The recovered dual-band signals may also suffer errors. Therefore, \name\ needs to identify these abnormal signals and improve them with more reliable ones.
First, abnormal detection based on hypothesis testing is conducted to identify abnormal signals. Then \name{} will try to recover them from the previously constructed database. 
As for fingerprinting for the first time, the robot just recollects fingerprints thereof to calibrate these signals.

\subsection{Detection of Abnormal Signal}
\name's identification of abnormal signals is grounded on the assumption that signal propagation is smooth,  which is justified by Eq.~\ref{eq:shadow_log}. So, the basic idea is to identify samples which obviously deviate from the smooth signal model. Specifically, \name's detection of abnormal signal includes two steps: (i) fitting the spatial signal model with a Gaussian process and estimating the norm value for each measured signal, and (ii) performing the largest residual test to identify abnormal signals.

\vspace{+3pt}
\noindent $\bullet$ {\bf Estimation with Gaussian Process.~}
The Gaussian process regression is used to fit the spatial distribution of the signal strength for two reasons. First, a Gaussian process assumes that measurements are drawn from random variables conforming Gaussian distributions, as analytically/empirically corroborated in Eq.~\ref{eq:shadow_log} and Fig.~\ref{fig:fit}. Second, the Gaussian process regression predicts the distribution of RSSI at a certain point, including the mean and varianc, which facilitates the next step of \name, i.e., the largest residue test. 
Let the fingerprint training set be a collection of fingerprints $\mathcal{D} = \{ ({x_1},{y_1}),({x_2},{y_2}), \ldots ,({x_n},{y_n})\}$,
where $x$ denotes 2-D coordinates and $y$ denotes the RSSI of an AP. It is assumed that the measured signal strength $y$ consists of a {\em true} signal strength $f(x)$ and an independent Gaussian noise $\omega \in N(0,{\sigma ^2})$, i.e., ${y_i} = f({x_i}) + {\omega _i}$. The collection of $f({x})$ is to be drawn from a Gaussian process, thus it conforms a multivariate Gaussian distribution with mean function $m(\cdot)$ and covariance function $k(\cdot, \cdot)$:
{
\footnotesize
{
\begin{equation}
\begin{bmatrix}
f(x_1)\\ 
\vdots \\ 
f(x_n)
\end{bmatrix}
\sim
{N}
\left (\begin{bmatrix}
m(x_1)\\ 
\vdots \\ 
m(x_n)
\end{bmatrix}  
,
\begin{bmatrix}
k(x_1, x_1) & \cdots  & k(x_1, x_n) \\ 
\vdots  & \ddots  & \vdots  \\ 
k(x_n, x_1) & \cdots  & k(x_n, x_n) 
\end{bmatrix}
\right ) .
\label{eq:GP_model}
\end{equation}
}

}

%
%
In general, mean function $m(\cdot)$ is set to $0$, and kernel functions are used to represent the covariance $k(\cdot,\cdot)$. Here the squared exponential kernel is adopted:
\begin{equation}
    k(x_i, x_j) = \sigma _f \exp\left ( -\frac{1}{2l^2}\left \| x_i-x_j \right \|^2 \right ),
    \label{eq:se_kernel}
\end{equation}
where $\sigma _f$ is signal variance and $l$ is a length scale. Both parameters determine the smoothness of the function $f(x)$ estimated by the Gaussian process.
%

Then we represent the distribution of $y=f(x)+w$. Since it has a zero-mean noise, its mean function is still $0$.
The noise terms can be incorporated into the covariance function:
\begin{equation}
    \mathrm{cov} ({y_i},{y_j}) = k(x_i, x_j) + ({\sigma ^2}){\delta _{ij}},
\end{equation}
where ${\delta _{ij}} = 1 $ if $i = j$ and zero otherwise. 

Denote the testing set as $\mathcal{T} = \{ ({x^*_1},{y^*_1}),({x^*_2},{y^*_2}), \ldots ,({x^*_m},{y^*_m})\}$, which is drawn from the same unknown distribution as $\mathcal{D}$. For notational convenience, we aggregate $n$ input vectors $x_i$ of $\mathcal{D}$ into $n \times 2$ matrix $X$, $n$ output values $y_i$ of $\mathcal{D}$ into $n \times 1$ vector $Y$, $m$ input vectors $x^*_i$ of $\mathcal{T}$ into $m \times 2$ matrix $X^*$, $m$ output values $y^*_i$ of $\mathcal{T}$ into $m \times 1$ vector $Y^*$. The training points and testing points must have a joint multivariate Gaussian distribution:
{
\small
\begin{equation}
\begin{bmatrix}
Y\\
Y^*
\end{bmatrix}
\sim 
N
\left( 0,
\begin{bmatrix}
K(X,X)+\sigma^2 I &  K(X,X^*) \\
K(X^*,X) & K(X^*,X^*) + \sigma^2 I
\end{bmatrix}
\right),
    \end{equation}
where 
\begin{equation}
\begin{aligned}
        K(X,X)[i,j]=k(x_i,x_j)&, K(X,X^*)[i,j]=k(x_i,x^*_j),\\ K(X^*,X)[i,j]=k(x^*_i,x_j)&, K(X^*,X^*)[i,j]=k(x^*_i,x^*_j).
\end{aligned}
\end{equation}
}

With the rules for conditional density, we get the predicted value at $X^*$ conditioned on training data $X, R$:
$Y^*|X^*,X,Y \sim \mathcal{N}({\mu ^*}, {\Sigma ^*})$, where

\begin{equation}
\begin{aligned}
{\mu ^*} = &K({X^*},X){(K(X,X) + \sigma^2 I)^{ - 1}}Y\\
\Sigma ^* = &K({X^*},{X^*})+\sigma ^2 I \\
&- K({X^*},X){(K(X,X) + \sigma^2 I)^{ - 1}}K(X,{X^*})    
\end{aligned}
\label{eq:GP_prediction}
\end{equation}

As can be seen from Eq.\ref{eq:GP_prediction}, the predicted mean is a linear combination of observed signal strengths $Y$, and the weights depends on covariance $K({X^*},X)$, while the squared exponential kernel determines that nearby function values are highly correlated. On the other words, it is believed that RSSIs are locally smooth, thus the data will be regraded as outliers if it disagrees with our prior knowledge. 
%

To examine the collected fingerprints, we predict signal strengths of the training data, i.e., $X*=X$. Thus, a measured value $y$ and its expectation and variance $\mu^*_x, \Sigma^*_x$ are obtained for each location $x$.

%
%
\name\ trains the parameters using scikit-learn~\cite{williams2006gaussian}.

\vspace{+3pt}
\noindent $\bullet$ {\bf Largest Normalized Residual Test.~}
The outlier identification is through the analysis of residues. Normalizing residues is necessary for us to find which one most deviate the estimation:
\begin{equation}
r^N = \frac{y-\mu^*_x}{\sqrt{\Sigma^*_x}} \sim \mathcal{N}(0,1).
\end{equation}

Then existence of outliers can be verified by the following hypothesis test:
\begin{itemize}
    \item if any $|r^N|>t$ in collected fingerprints, there is a suspicion of bad data.
    \item if all $|r^N| \leq t$, the hypothesis that there is no bad data is supported.
\end{itemize}
\name\ set $t$ to 1.96.
%

From \cite{grubbs1950sample}, it is shown that for a measurement set the measurement with the largest normalized residual contains a gross error. As a result, one abnormal signal can be identified by testing $r^N_{max} > t$. To identify all abnormal signal, the largest normal residual test performs within a loop: 
\begin{enumerate}
	\item Estimate expectations and variances from the training set.
	\item If the largest residue exceeds the threshold, withdraw it from the training set, then go to step one. If not, finish the test. 
\end{enumerate}
%

\subsection{Signal Recovery}
\begin{figure}[htbp]
    \centering
    \includegraphics[width=0.98\linewidth]{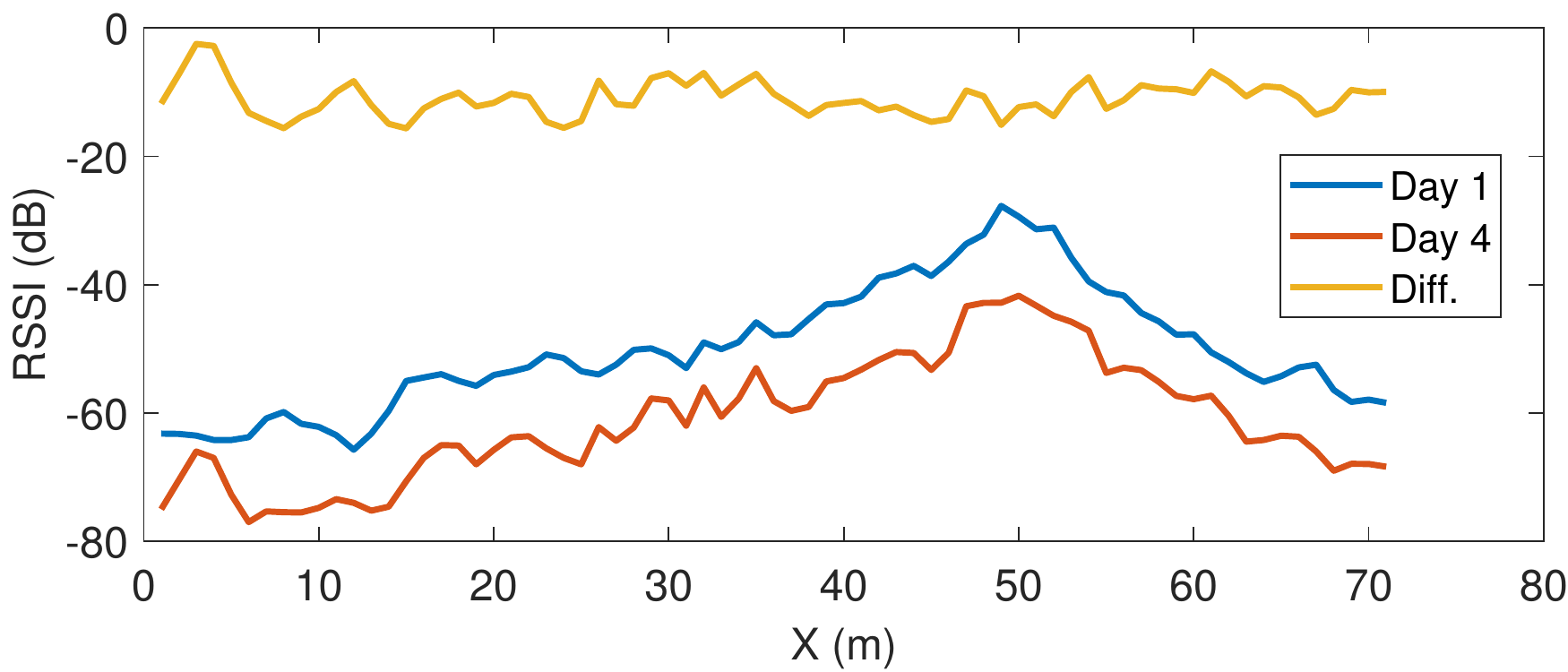}
    \caption{Difference of the signal strength in three days.}
    \label{fig:shift}
\end{figure}

The next step is to recover these abnormal signals from the previously constructed fingerprint database.  
Although the pattern of WiFi change after a long time is difficult to analyze~\cite{7174948}, the short time change of the WiFi signal is (relatively) predictable. We collect samples along a path in three days with the same collection method in Fig.~\ref{fig:24and5GHz}. The alternation of the signal after three days can be approximated as a shift, as shown in Fig.~\ref{fig:shift}. So we shift the corresponding signals in the past database to recover the abnormal signals.

\section{Evaluation}
\label{sec:evaluation}

We present our evaluation of \name\ in this section. First we explain the experiment settings in Section~\ref{sec:sec:methodlogy}. Then the time and energy efficiency of \name{} is compared to the baseline in Section.~\ref{sec:sec:time_energy}. In Section.~\ref{sec:sec:signal_on_fingerprintmap} the impacts of the signal recovery methods on the spatial domain and on the time domain on the fingerprint database are visualized. Finally, the localization accuracy of \name{} is evaluated in Section.~\ref{sec:sec:accuracy}.

\begin{figure}[htbp]  
  \centering
  \begin{minipage}[b]{0.4\linewidth}
  \begin{subfigure}[t]{1\linewidth}
  \centering
  \includegraphics[width=1\linewidth]{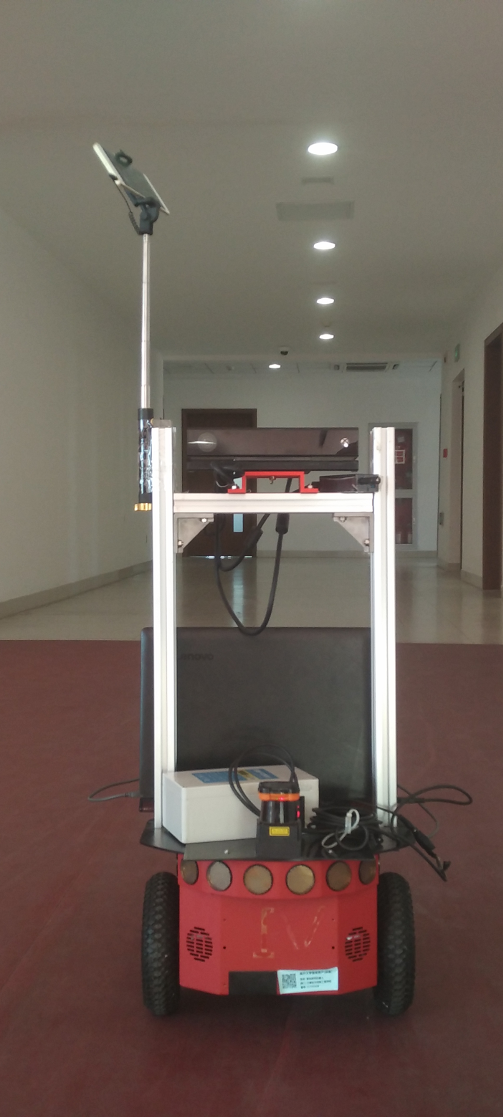}
  \caption{}
  \end{subfigure}
  \end{minipage}
  \begin{minipage}[b]{0.58\linewidth}
  \begin{subfigure}[t]{1\linewidth}
  \centering
  \includegraphics[width=1\linewidth]{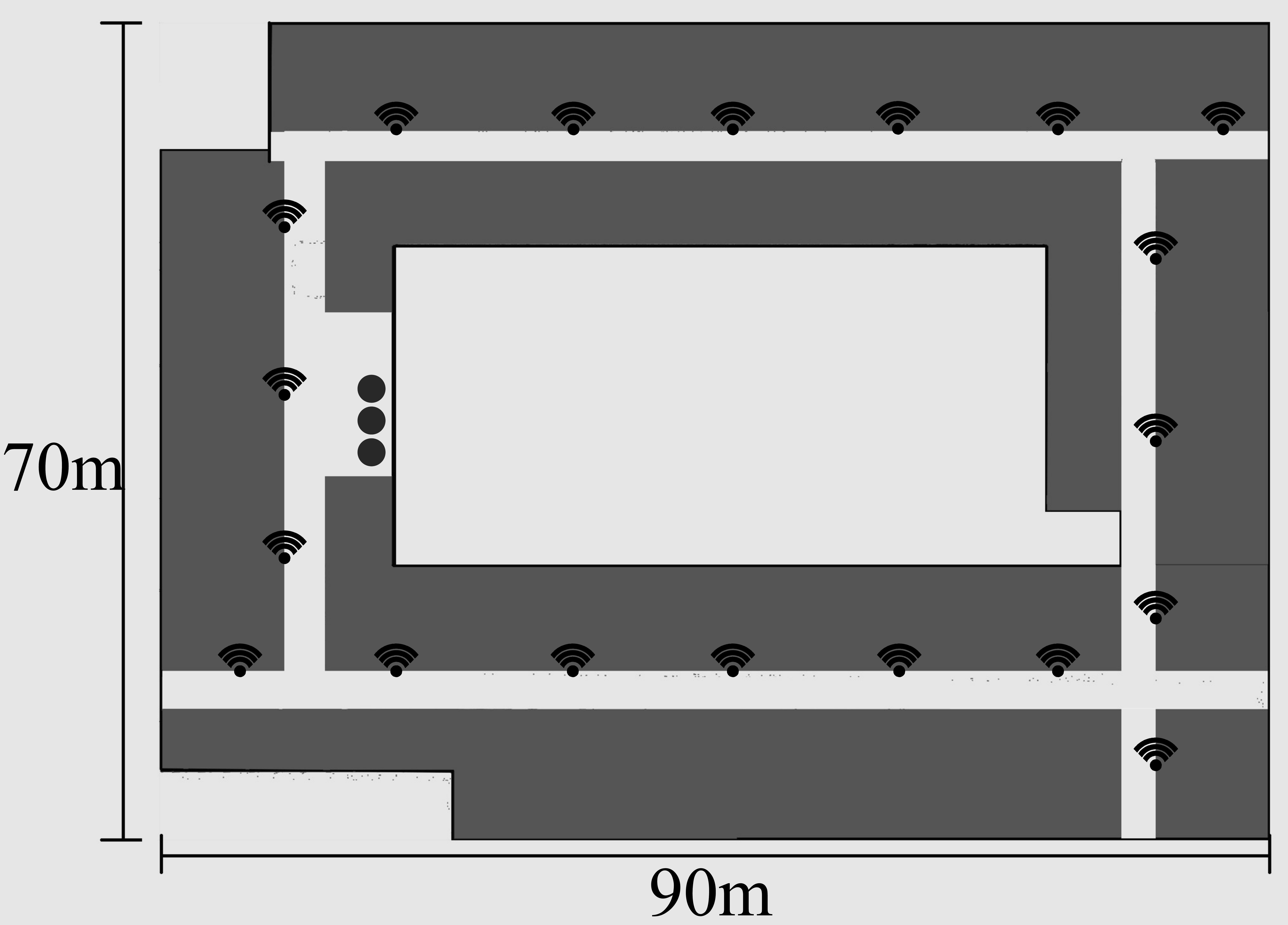}
  \caption{}
  \end{subfigure}
  \\
  \begin{subfigure}[t]{1\linewidth}
  \centering
  \includegraphics[width=1\linewidth]{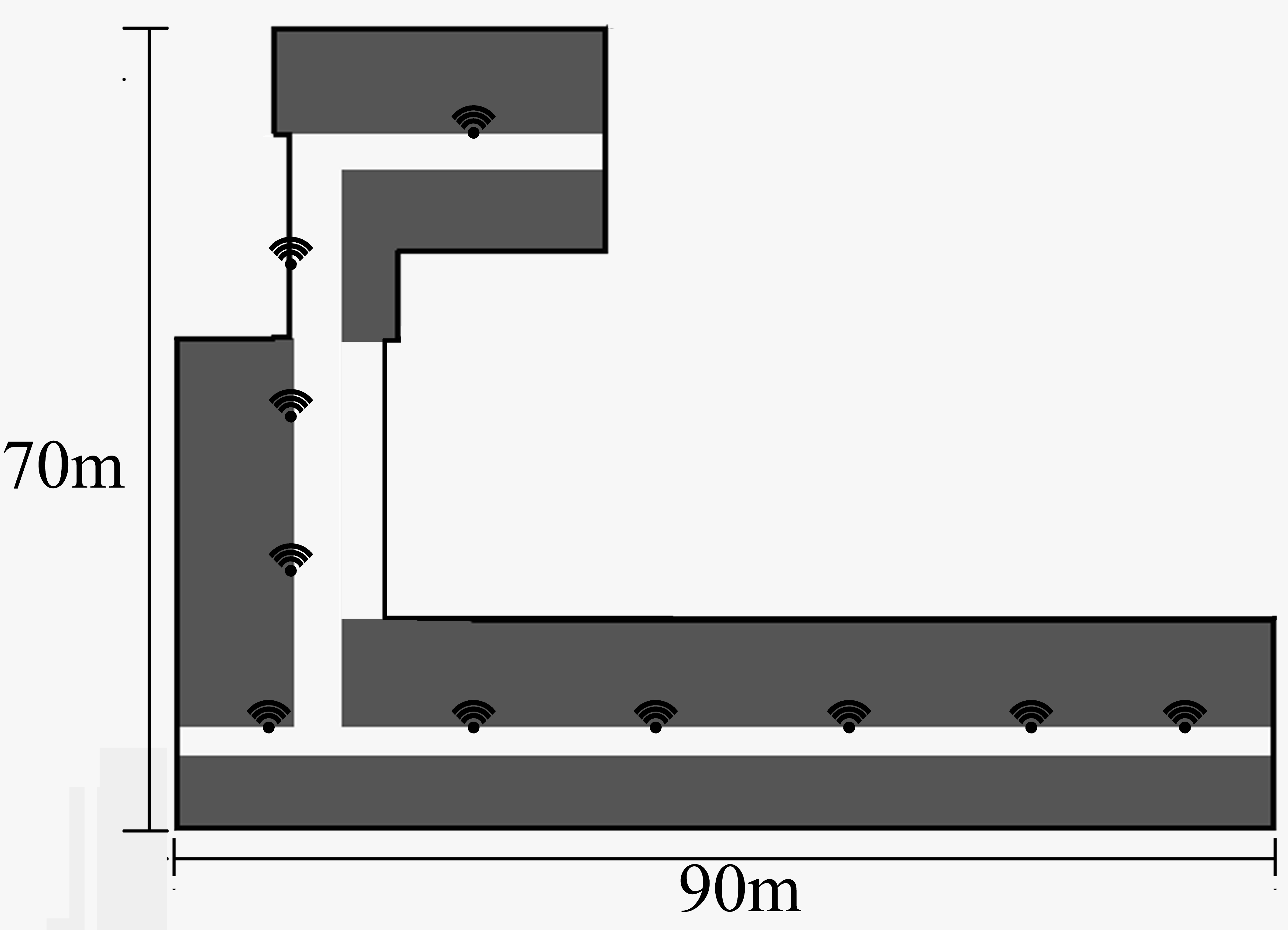}
  \caption{}
  \end{subfigure}
  \end{minipage}
  \caption{(a) SLAM-enabled robot. (b) (c) are separately 3rd and 6th floors of our Department building.}
  \label{fig:robot}
\end{figure}

\subsection{Methodology}
\label{sec:sec:methodlogy}
\vspace{+3pt}
\noindent $\bullet$ {\bf Experiment Settings.~}
We have deployed and evaluated \name{} on two sites (the 3rd floor and the 6th floor) of our Department building. These two areas and AP deployment there are shown in Figs.~\ref{fig:robot}(b)(c).
%
%
We use a Pioneer-3DX robot in Fig.~\ref{fig:robot} equipped with a HOKUYO UTM-30LX laser module as the agent. A LENOVO ideapad Y700 laptop with Ubuntu 16.04 operating system serves as the upper computer, and ROS (a robot operating system) is adopted to control the robot. A MI Note Pro smartphone is attached to the robot and used as the WiFi scanner during the site survey with 3 scan interval.
The robot surveys the floor at a speed of $0.5$m/s.
Before and after the site survey, we measure the voltage of the robot, the laser and the laptop. Then we use the discharge curves to calculate consumed power.
We update the fingerprint database every three days in twelve days.

\vspace{+3pt}
\noindent $\bullet$ {\bf Baseline.~}
For comparison, we also implement the survey-with-sojourn method: The robot surveys the building according to the same route as \name{}'s survey, but sojourns at each grid point generated in Sec.~\ref{sec:floor_recognition} for 10s. 




\vspace{+3pt}
\noindent $\bullet$ {\bf Gaussian Process Fingerprint Map.~}

For visualization of the fingerprint database and localization, Gaussian process fingerprint maps are separately generated using the fingerprint database of \name{} and the database of the baseline.
Specifically, a Gaussian process model is trained by fingerprints after completing the site survey, which outputs the mean and variance of the signals' RSSI at every grid point generated in Sec.~\ref{sec:floor_recognition}. \name\ uses these RSSI statistics as the fingerprint thereof. 
%


\vspace{+3pt}
\noindent $\bullet$ {\bf Localization Method.~}

We implemented the following simple but classic localization methods on top of the fingerprint map constructed above, and examined the resultant localization accuracy when a person holding a smartphone walks around on the two sites.

\vspace{+3pt}
{\em \underline{(1)~~Bayes.}~}
The first localization method exploits the fingerprint map constructed by \name\ with Bayes method~\cite{roos2002probabilistic}.
Denote the reference locations as $L = \{ {l_1},{l_2}, \ldots ,{l_m}\}$, and the observation vector as $o_{1\times b}$.
The reference location with the maximum probability $p(l|o)$ is used as the predicted location $\hat l$:
\begin{equation}
\begin{aligned}
&p(l|o) = {\frac{{p(o|l)p(l)}}{{p(o)}}}\propto p(o|l), 
\\
&\hat l = \mathop {\arg \max }\limits_{{l_j}\in{L}} p(l_j|o)=\mathop {\arg \max }\limits_{{l_j}\in{L}}\prod\limits_{i = 1}^b {G({o_i}|{l_j})} .
\end{aligned}
\end{equation}
where $G$ is the probability of $v_i$ in the corresponding Gaussian distribution. 
We then further improve the thus-obtained location results with particle filter.\footnote{Please see \cite{evennou2006advanced} for details of particle filter.}


\vspace{+3pt}
{\em \underline{(2)~~KNN.}~}
We also implemented a KNN-based localization method on top of \name, i.e., locating the online collected WiFi signals to the $K$ reference locations with the closest WiFi fingerprints. We used a K of $2$ unless specified otherwise. Again, particle filter is then used to further improve the localization accuracy.





\begin{table}[htbp]
\caption{Time and energy cost to survey 3F.}
\centering
\begin{tabular}{ccccc}
\hline
Method                   & Time  & Robot & Laser  & Laptop  \\ \hline
\name{} & 43 min       & 30 Wh   & 7 Wh     & 23 Wh    \\
Baseline                 & 121 min      & 69 Wh  & 17 Wh   & 67 Wh    \\ \hline
\end{tabular}
\label{table:time_3F}
\end{table}

\begin{table}[htbp]
\caption{Time and energy cost to survey 6F.}
\centering
\begin{tabular}{ccccc}
\hline
Method                   & Time  & Robot  & Laser  & Laptop  \\ \hline
\name{} & 51 min       & 33 Wh   & 8 Wh    & 41 Wh    \\
Baseline                 & 177 min      & 90 Wh   & 25 Wh   & 113 Wh   \\ \hline
\end{tabular}
\label{table:time_6F}
\end{table}

\subsection{Time Overhead of Site Surveys}
\label{sec:sec:time_energy}
Tables.~\ref{table:time_3F} and \ref{table:time_6F} summarize the time and energy overhead to survey the sites by \name{} and the baseline.  
Clearly, travel-without-sojourn requires far less time than travel-with-sojourn: \name{} saves 64\% time on the 3rd floor when compared to the traditional survey method, while a 71\% reduction is achieved on the 6th floor. 
We also compare the energy overhead. On 3F, the energy consumed by the robot motors is reduced by 57\% with \name{} compared to the baseline. The de/acceleration is quite frequent when survey the building with sojourn, due to the dense survey locations required for accurately modeling the spatial distribution of signal strength. In our experiments, the robot performs de/acceleration once for every 0.8m on average. 
\name{} also saves 59\% energy of the laser on 3F, which operates at constant power. 
We can see that the laptop is a major part of energy consumption, since it performs intense computation to localize and navigate the robot. On 3F, \name{} consumes 66\% energy less than the baseline. 
In total, \name{} reduces the energy consumption by 61\%.

\subsection{Signal Recovery on Fingerprint Map}
\label{sec:sec:signal_on_fingerprintmap}
\begin{figure}[htbp]
    \centering
    \begin{subfigure}[t]{0.98\linewidth}
    \centering
        \includegraphics[width = 1 \linewidth]{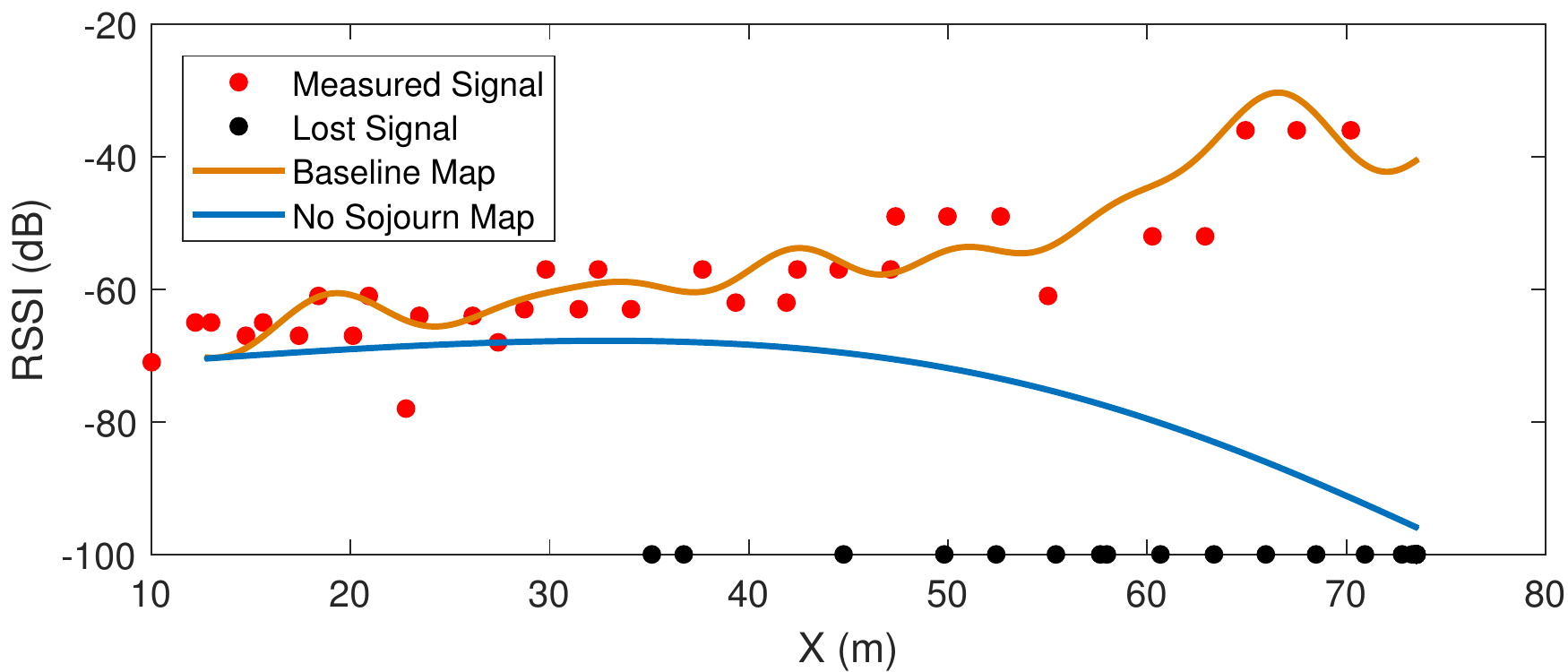}
        \caption{}
    \end{subfigure}
    \\
    \begin{subfigure}[t]{0.98\linewidth}
    \centering
        \includegraphics[width = 1 \linewidth]{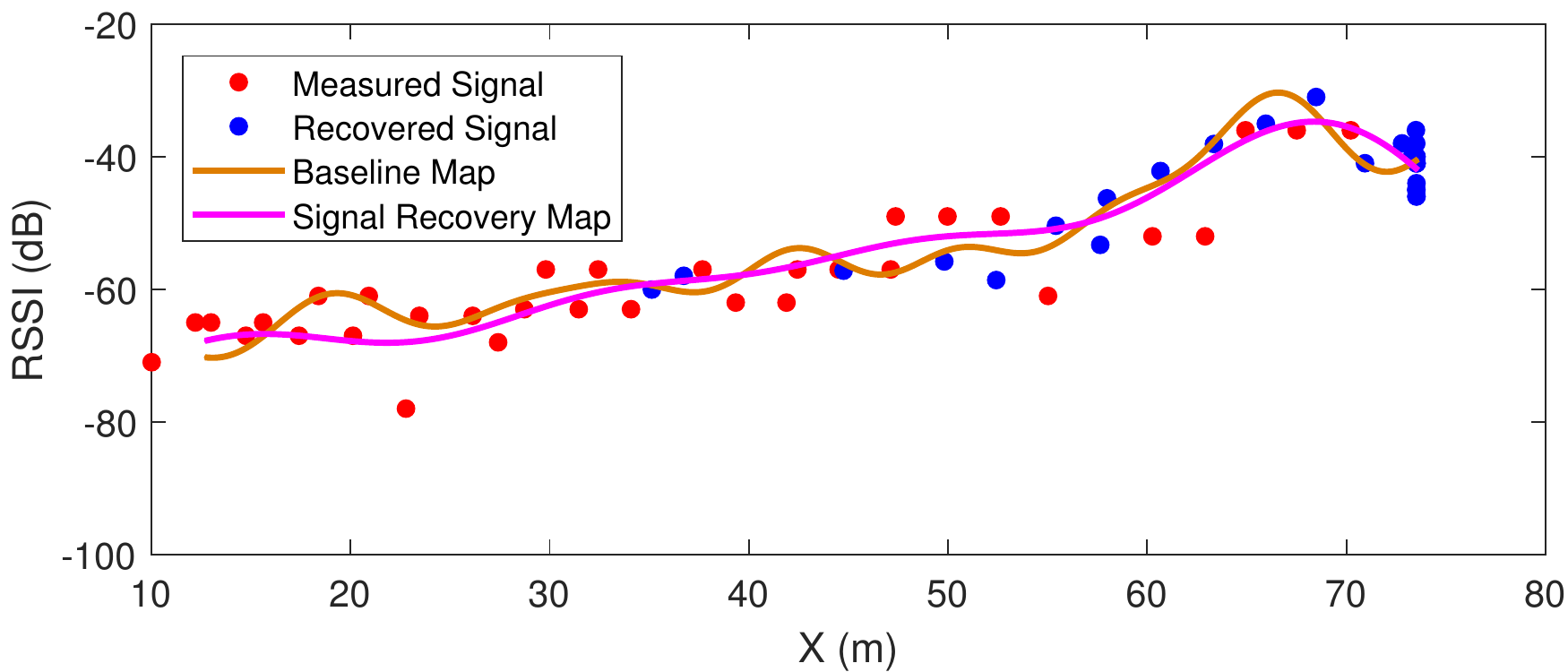}
        \caption{}
    \end{subfigure}
\caption{The fingerprint map constructed w/o and w/ lost signal recovery.}
\label{fig:GP_raw_recovered_model}
\end{figure}

Next we examine the impact of lost signal recovery on the fingerprint map.
Fig.~\ref{fig:GP_raw_recovered_model} shows the fingerprint maps constructed by \name{} with and without signal recovery using the dual-band signals, in which {\em Measured Signal} denotes fingerprints with the measured signal, {\em Lost Signal} denotes fingerprints with the lost signal, which is set to -100 dBm, and {\em Recovered Signal} denotes fingerprints with the recovered signal.
{\em No Sojourn Map} in Fig.~\ref{fig:GP_raw_recovered_model} represents the mean of the fingerprint map constructed with only data collected during survey without sojourn, and {\em Signal Recovery Map} represents the mean of the fingerprint map constructed with lost signal recovery, while {\em Baseline Fingerprint map} represents the mean of the fingerprint map constructed with data of the baseline.
Signal loss occurs even when strong signals could be received, causing significant uncertainty in the thus-constructed fingerprint map.
Such uncertainty can be effectively mitigated via \name's signal recovery, as observed in Fig.~\ref{fig:GP_raw_recovered_model}.

\begin{figure}[htbp]
    \centering
    \begin{subfigure}[t]{0.98\linewidth}
    \centering
        \includegraphics[width = 1 \linewidth]{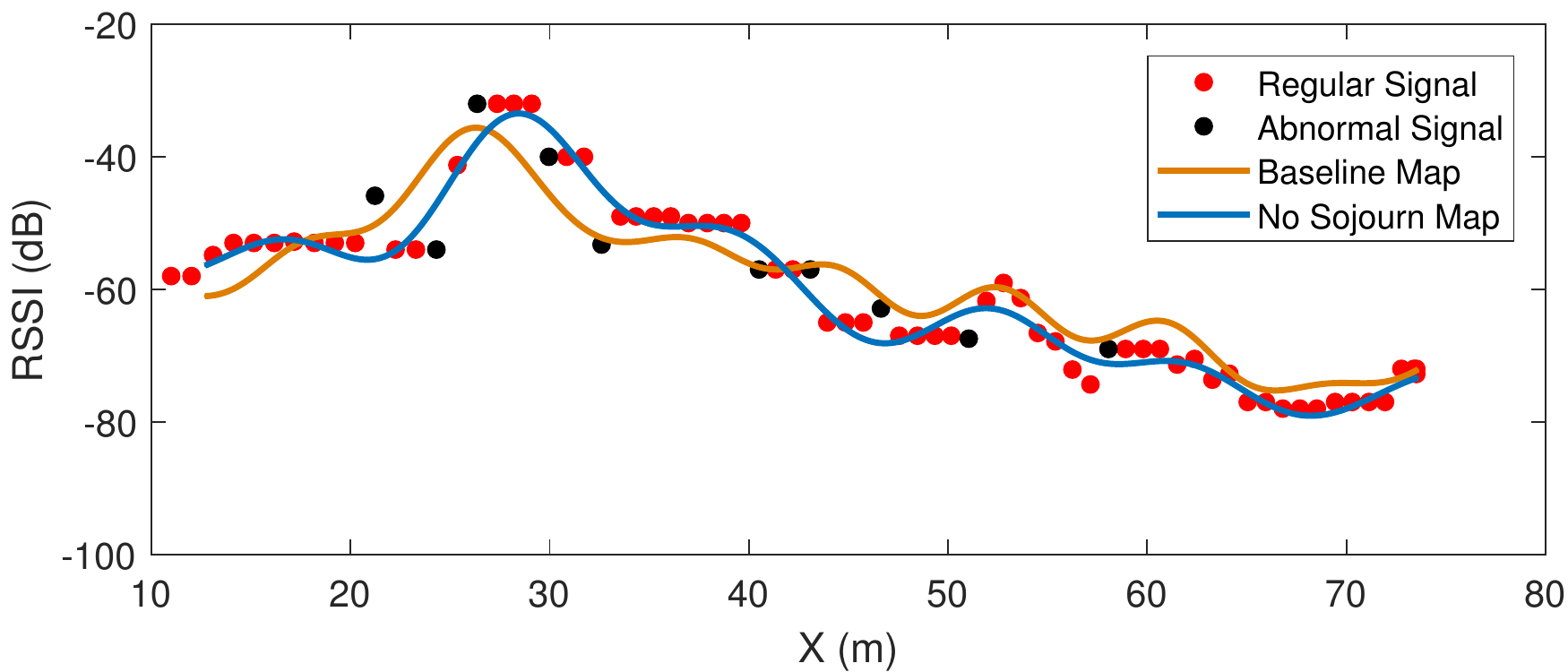}
        \caption{}
    \end{subfigure}
    \\
    \begin{subfigure}[t]{0.98\linewidth}
    \centering
        \includegraphics[width = 1 \linewidth]{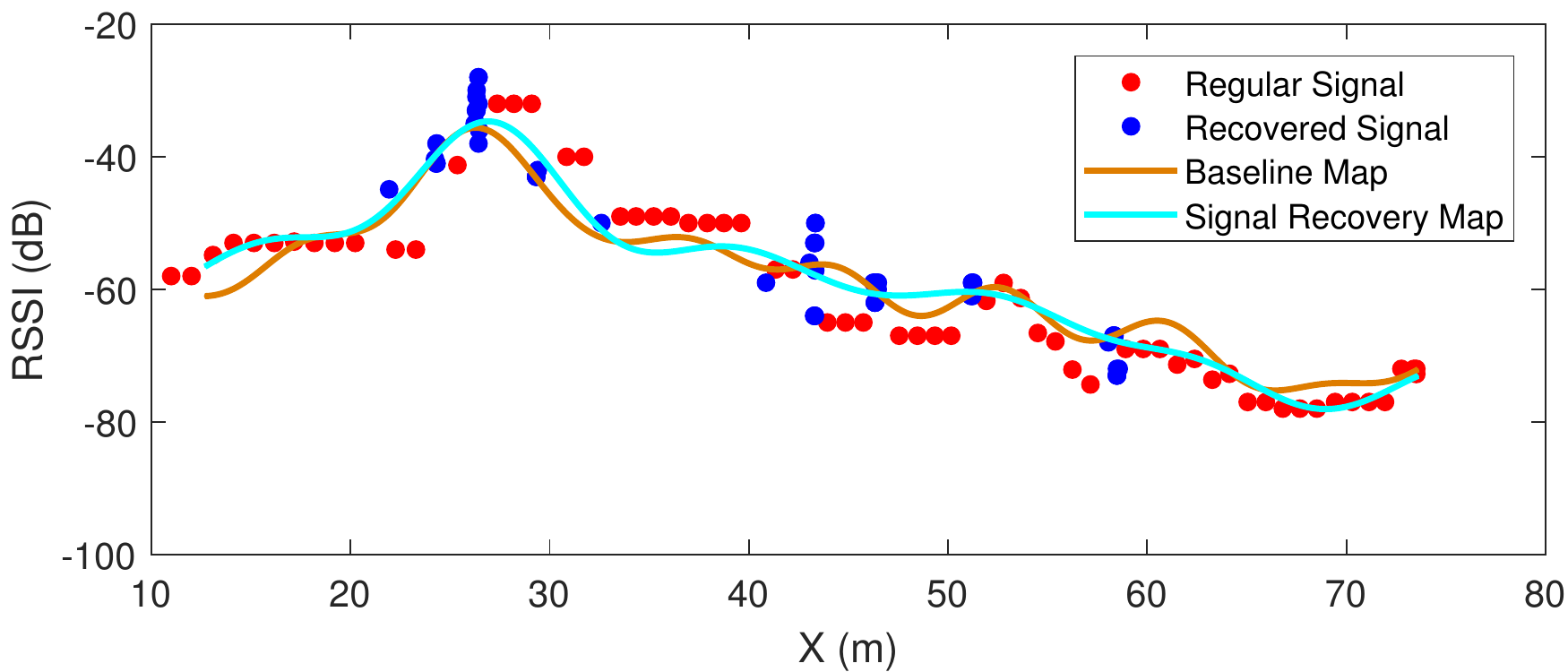}
        \caption{}
    \end{subfigure}
\caption{Localization w/o and w/ abnormal signal recovery.}
\label{fig:resurvey_fg}
\end{figure}
We next examine the impact of abnormal signal recovery on the fingerprint map.
Fig.~\ref{fig:resurvey_fg} shows fingerprint maps constructed with and without abnormal signal recovery, in which {\em Regular Signal} denotes fingerprints with the regular signal, {\em Abnormal Signal} denotes fingerprints with the identified abnormal signal, and {\em Recovered Signal} denotes fingerprints with signals recovered by previous data.
{\em No Sojourn map} and {\em Baseline Fingerprint map} in Fig.~\ref{fig:resurvey_fg} have the same meaning as in Fig.~\ref{fig:GP_raw_recovered_model}, while {\em Signal Recovery map} represents the mean of fingerprint map constructed with abnormal signal recovery.
As can be seen from Fig.~\ref{fig:resurvey_fg}, non-smooth points are identified as abnormal signals. These signals make the local shapes of the fingerprint map deviate from the ground truth. But abnormal signal recovery calibrates these flaws and make the fingerprint map of \name{} almost coincide with the baseline map. 
%

 
%


\subsection{Localization Accuracy}
\label{sec:sec:accuracy}
Next we examine \name{}'s impact on localization accuracy. The examination is three-fold: (i) examine the impact of signal recovery; (ii) examine the impact of signal completion; (iii) examine the impact of \name{} and compare the localization accuracy of \name{} with that of the baseline.


\noindent $\bullet$ {\bf Impact of Lost Signal Recovery.~}
\begin{figure}[htbp]
    \centering
    \begin{subfigure}[t]{0.48\linewidth}
    \centering
    \includegraphics[width = 1\linewidth]{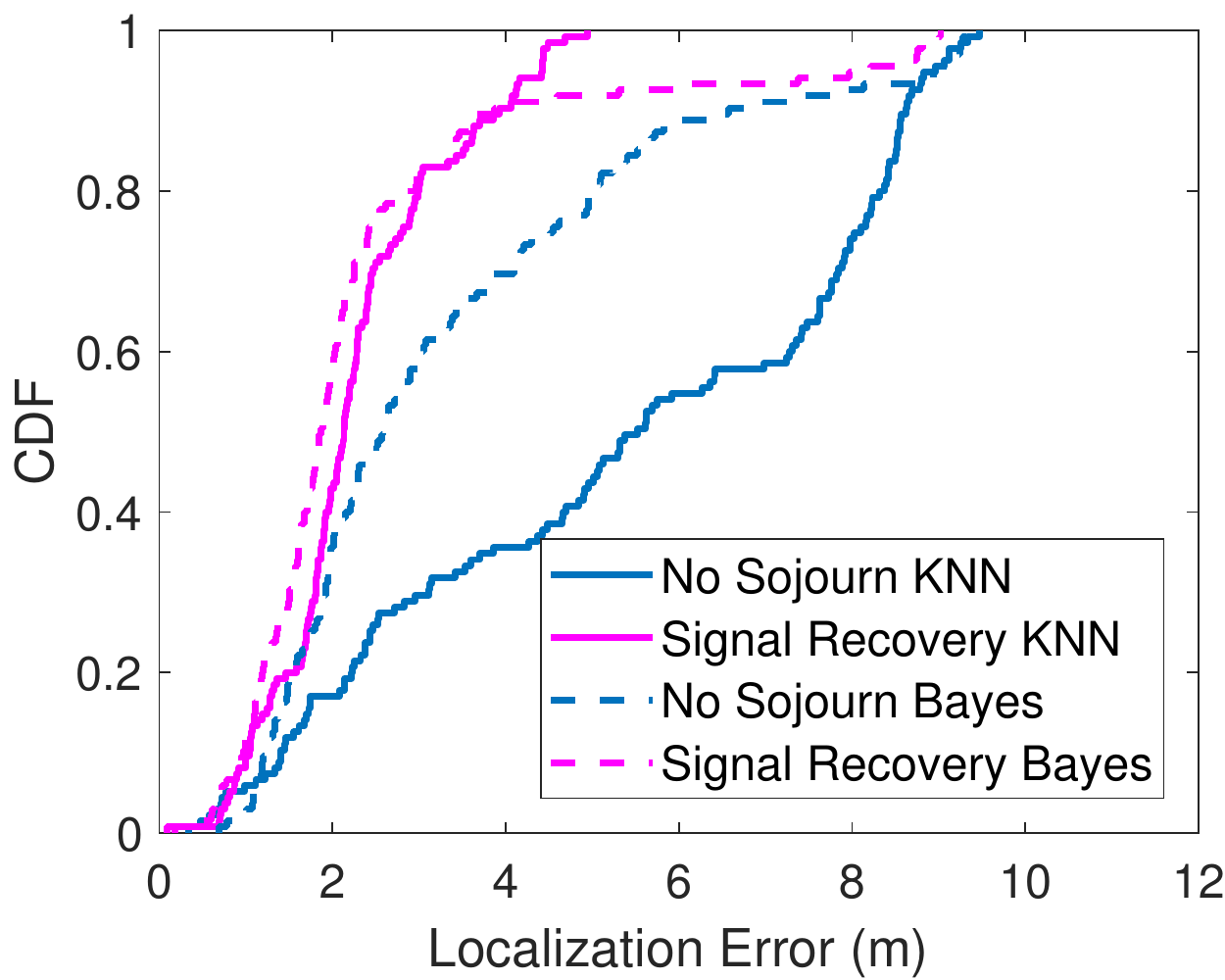}
    \caption{}
    \end{subfigure}
    \hfill
    \begin{subfigure}[t]{0.48\linewidth}
    \centering
    \includegraphics[width = 1\linewidth]{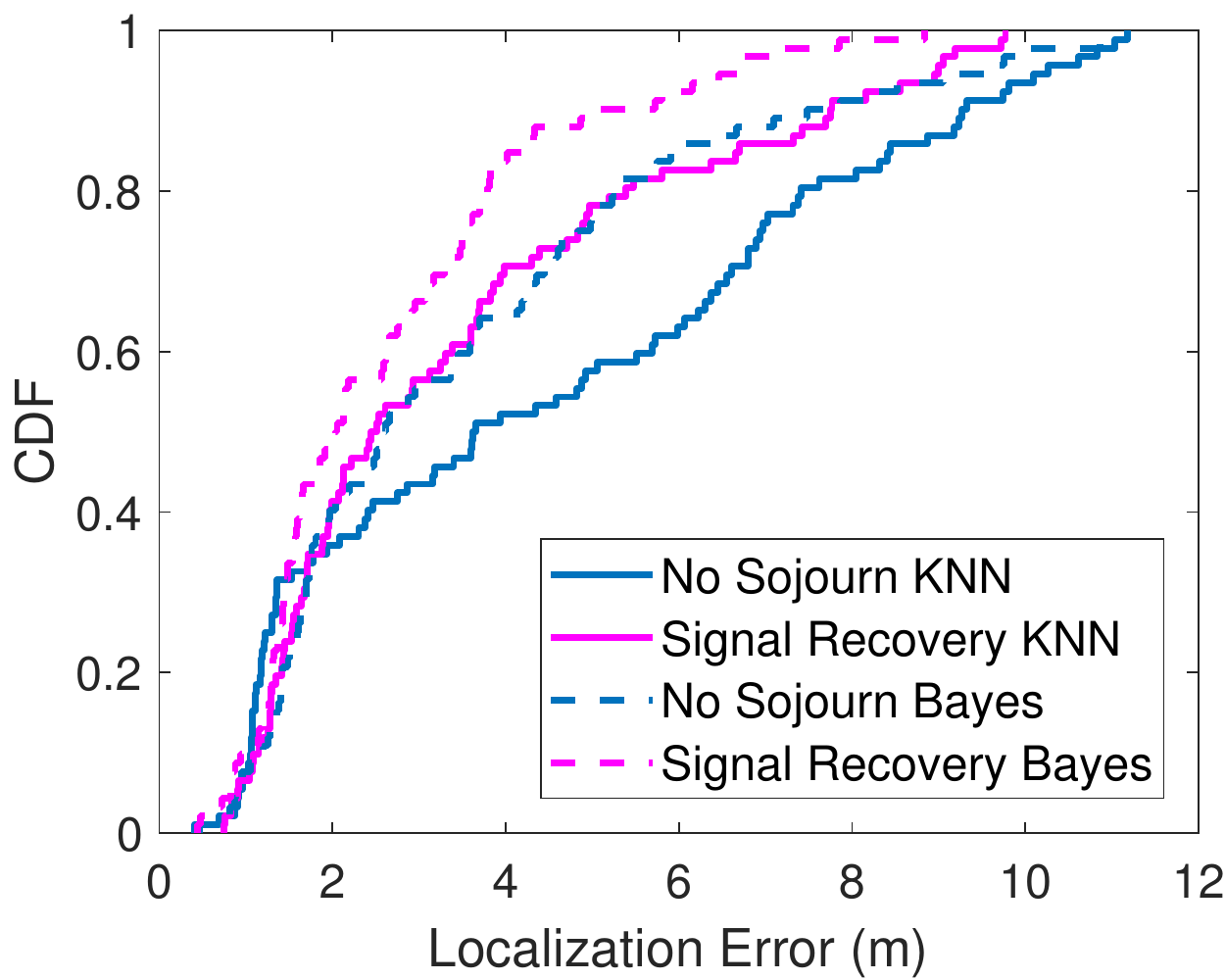}
    \caption{}
    \end{subfigure}
    \caption{localization w/o and w/ Lost signal recovery.}
    \label{fig:recovery_localization}
\end{figure}

Fig.~\ref{fig:recovery_localization} summarizes the localization results of fingerprint maps constructed with and without signal recovery, where {\em No Sojourn KNN} indicates using data collected during survey without sojourn and KNN method for localization, and the other legends' meaning is interpreted in the same way. In Fig.~\ref{fig:recovery_localization}(a) signal recovery improves mean error of KNN method from 5.4m to 2.3m. And localization in the narrow corridor is more difficult, but signal recovery still works. Seen from Fig.~\ref{fig:recovery_localization}, using Bayes method, mean error decrease from 4.5m to 3.5m.

But the max error does not degrades much with the help of lost signal recovery, e.g. max errors with/without signal recovery are close In Fig.~\ref{fig:recovery_localization}. This can be explained by the error caused by recovery from an abnormal signal, which is solved by abnormal signal recovery.        

\noindent $\bullet$ {\bf Impact of abnormal signal recovery.~}

\begin{figure}[t]
    \centering
    \begin{subfigure}[t]{0.48\linewidth}
    \centering
    \includegraphics[width = 1\linewidth]{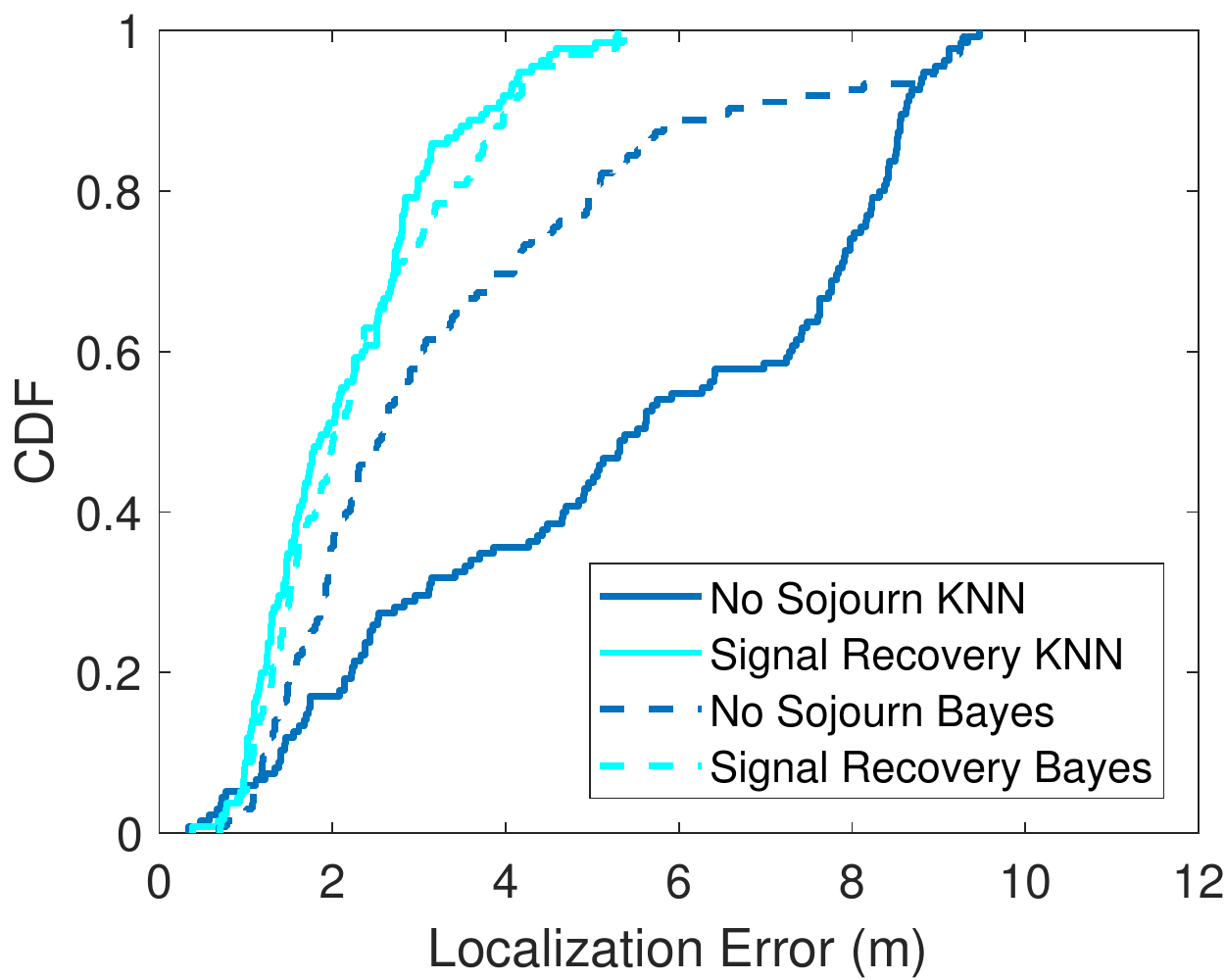}
    \caption{}
    \end{subfigure}
    \hfill
    \begin{subfigure}[t]{0.48\linewidth}
    \centering
    \includegraphics[width = 1\linewidth]{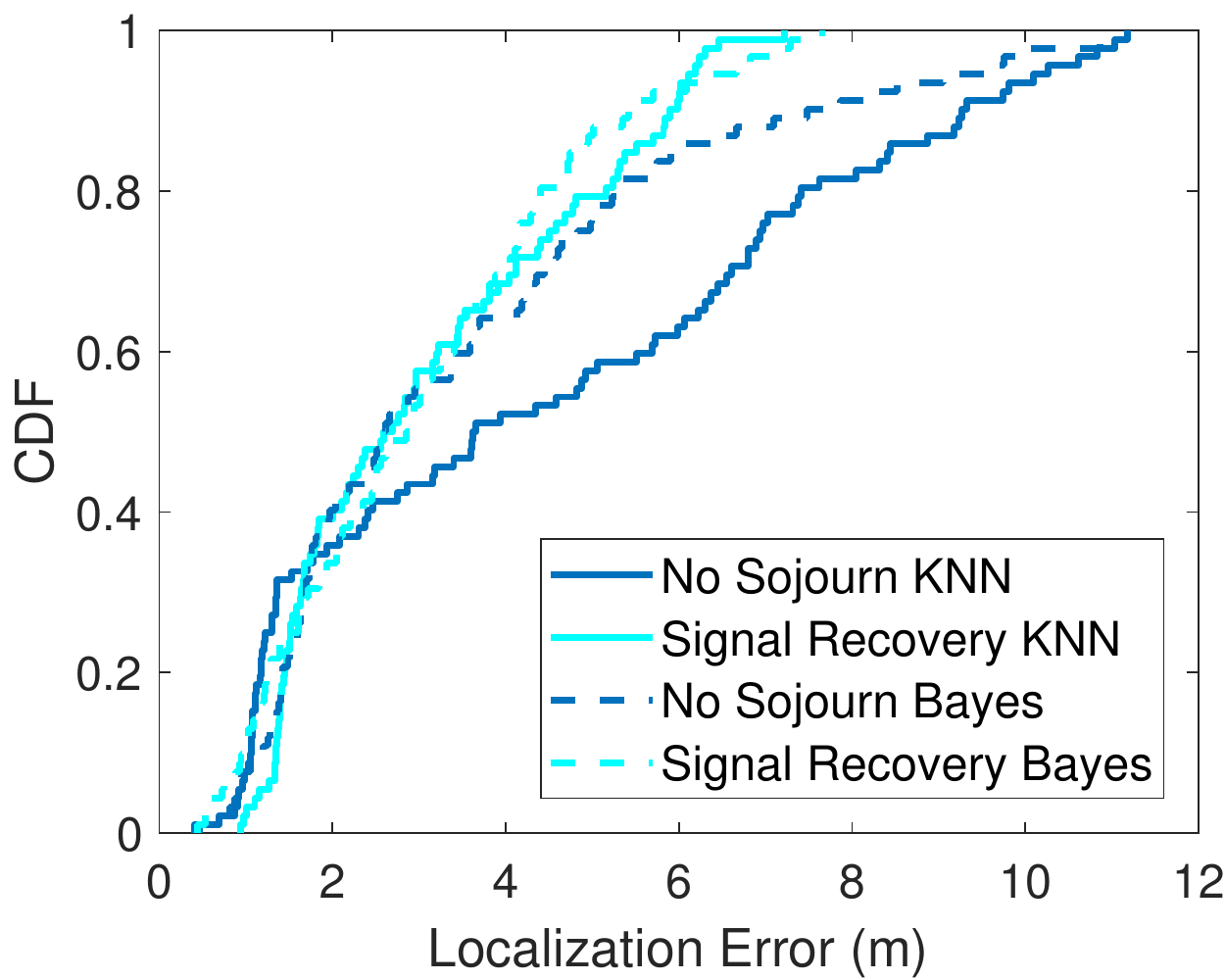}
    \caption{}
    \end{subfigure}
    \caption{Abnormal signal recovery's impact on localization.}
    \label{fig:supplement_localization}
\end{figure}

Fig.~\ref{fig:supplement_localization} summarize the localization results of fingerprint maps constructed with and without signal completion. Obviously both the mean error and the max error are depressed by signal completion. Take bayes method in Fig.~\ref{fig:supplement_localization}(a) as an example, the mean error decreases from 3.4m to 2.3m, while the max error decreases from 9.3m to 5.4m.

\noindent $\bullet$ {\bf Comparison with Baseline.~}
\begin{figure}[t]
    \centering
    \begin{subfigure}[t]{0.48\linewidth}
    \centering
    \includegraphics[width = 1\linewidth]{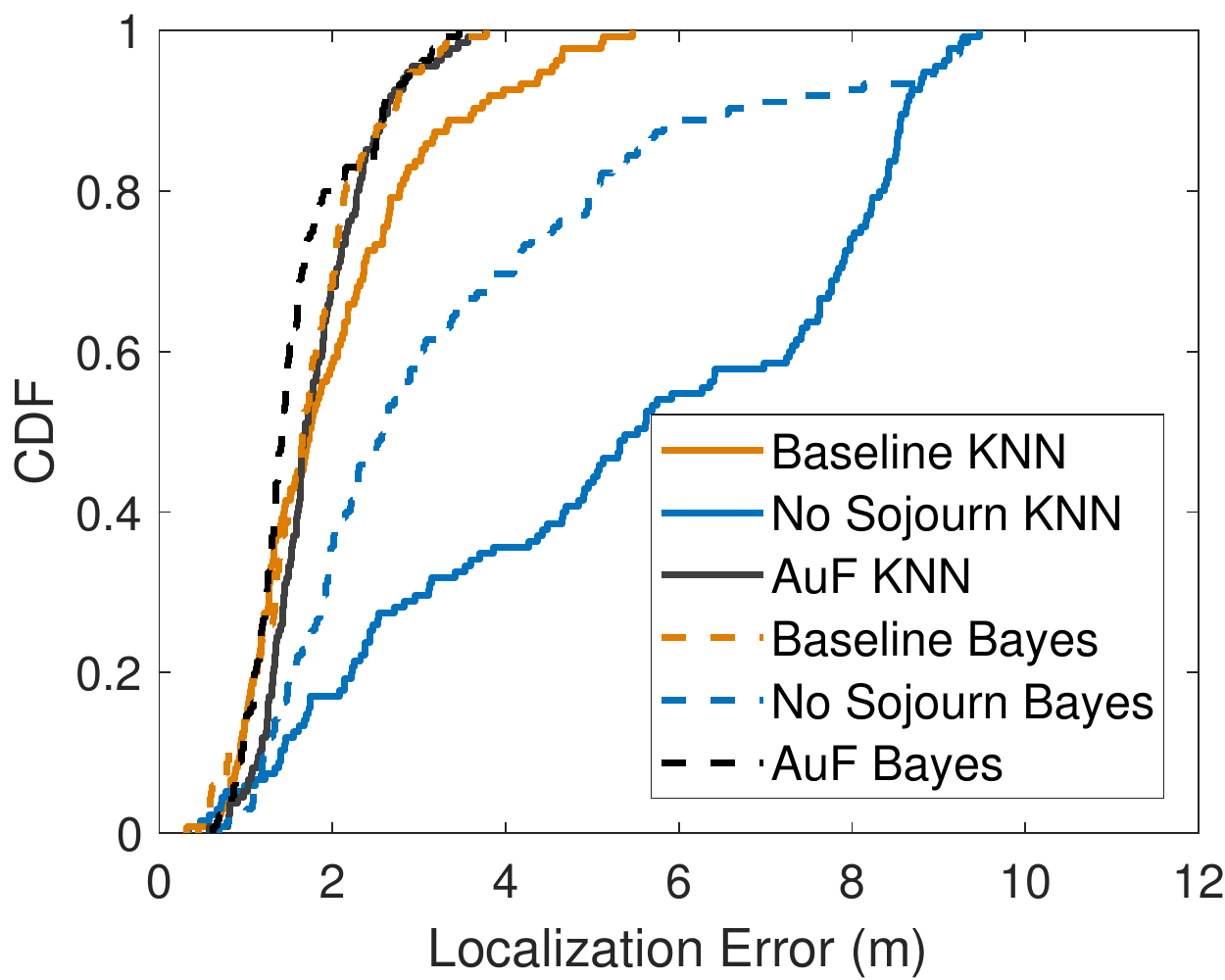}
    \caption{}
    \end{subfigure}
    \hfill
    \begin{subfigure}[t]{0.48\linewidth}
    \centering
    \includegraphics[width = 1\linewidth]{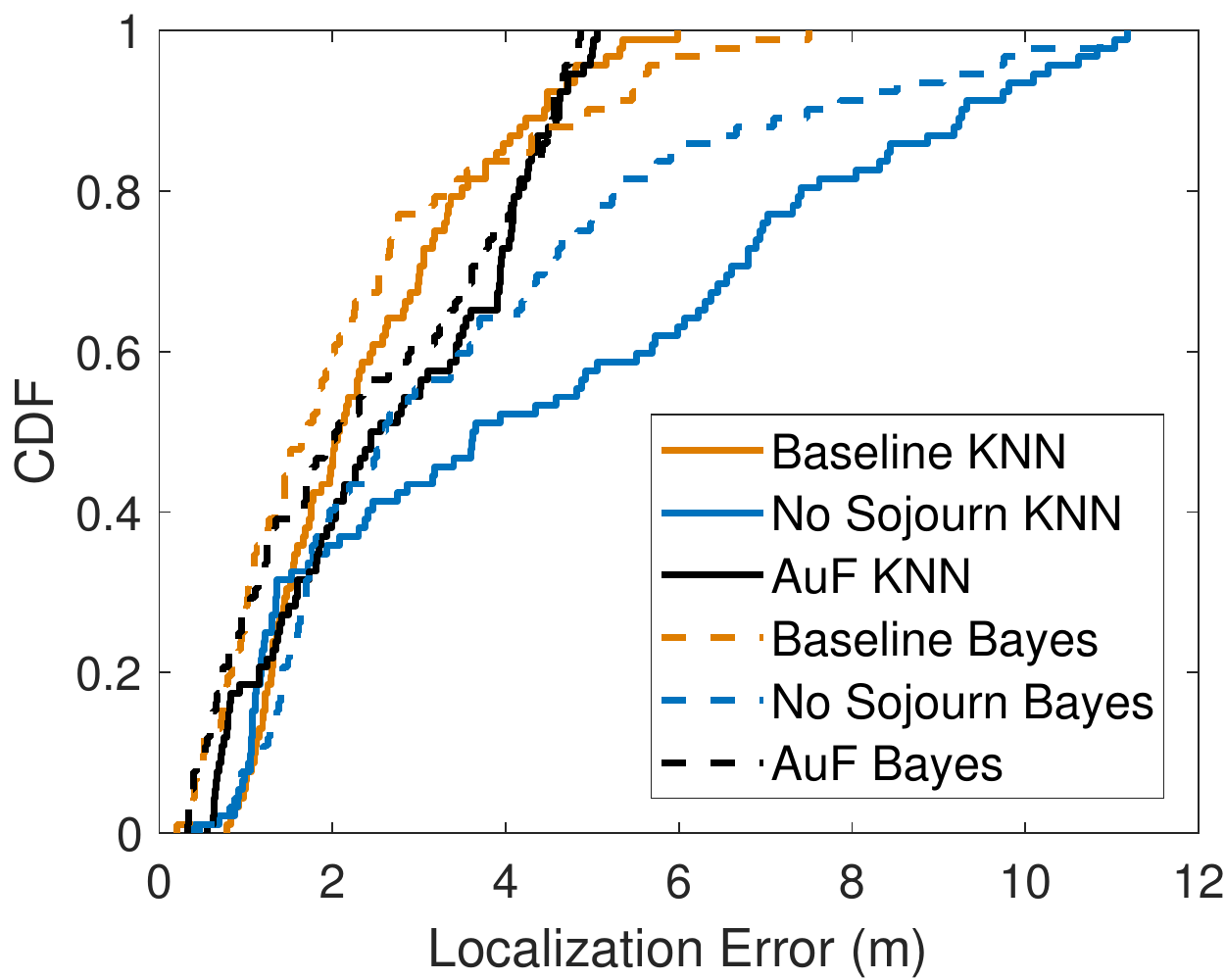}
    \caption{}
    \end{subfigure}
    \caption{Accuracy comparison between \name{} and the baseline.}
    \label{fig:compare_baseline_localization}
\end{figure}

The performance comparison in Fig.~\ref{fig:compare_baseline_localization} clearly shows that only improving the survey speed to reduce offline work is not desirable with the severely diminishing localization performance. In Fig.~\ref{fig:compare_baseline_localization}(a), using bayes method the mean error and the max error of raw increase 1.7m and 5.5m compared to the baseline.
%
On the other hand, the dataset provided by \name{} has no weakness in performance compared to the baseline. A slight improvement can be seen in Fig.~\ref{fig:compare_baseline_localization}(a), the point below which there are 80\% errors shifts from 2.1m to 1.9m, when adopts the fingerprint of \name{} rather than the baseline. In Fig.~\ref{fig:compare_baseline_localization}(d), \name{}'s has the little worse mean error but the little better max error. Using bayes method the baseline's mean error and max error is 2.2m and 7.5m, while \name{}'s mean error and max error is 2.4m and 4.9m. 

\section{Related Work}
\label{sec:related_work}
    Fingerprint-based localization has been extensively explored, and achieves fine-grained accuracy. The localization algorithms can be divided into two types: deterministic and probabilistic algorithms. Deterministic algorithms represent the signal strength as a scalar at a location. For example, RADAR~\cite{832252} takes nearest neighbor method to search the user's location from the database. Probabilistic algorithms establish distributions of signal strengths in database. Horus~\cite{youssef2005horus} is a representative instance, which uses a Bayesian network model. Furthermore, sensor  fusion~\cite{priyantha2005mobile} is studied to achieve further improved  localization accuracy. \name{}, as a fingerprint collection system, can be deployed to support all these localization methods.
    
    Also researches are carried out for the overhead of fingerprint map construction. 
    Crowdsourcing the signals from the users has also attracted much attention. 
    The early work, OIL~\cite{Park:2010:GOI:1814433.1814461} and Mole\cite{doi:10.1080/17489725.2012.692617} are designed to get fingerprints from users, but users are required to explicitly label his or her location for the collected signals.
    To solve human labelling problem, Unloc~\cite{wang2012no} and WiFi-SLAM~\cite{ferris2007wifi} combine dead-reckoning and WiFi signal patterns to localize walking users, and frees users from labelling their ground truth. 
    But the high dependence on the inertial sensor and the assumed walking pattern reduces accuracy of the fingerprint map.
     

     %
     
     
     Admittedly, in scenarios where only rough locations are needed, radio model-based approach and crowdsourcing approach are both convenient. 
     In contrast, \name{} can quickly construct fingerprint database while maintaining the localization performance.
     
     Using robots as professional surveyors has clear advantages. The robots free the human labor, and carry multiple devices to survey the floor while precise ground truth can be provided with a laser~\cite{lingemann2005high}, a depth camera~\cite{biswas2012depth} or just some sonars~\cite{varveropoulos2005robot}.
     
     The authors of \cite{mirowski2012depth,nguyen2016low} describe a process of WiFi mapping using an autonomous robot. However, they just simply make the robot survey with sojourn, thus the site survey is still time-consuming. More importantly, The robot's power consumption is non-negligible, rendering the deployment of the robot surveyor on the large buildings unacceptable. \name{} uses a more efficient survey method, facilitating its deployment in large space.

\section{Discussion: Computational Cost}
\label{sec:discussion}

A potential problem for \name{} is its computational cost from the signal recovery, especially the Gaussian process regression training process. The iterative training method's complexity with $N$ samples is $\mathcal{O}(N^2)$. Suppose that the number of abnormal signals is proportional to the number of samples, the computation complexity of \name{} is $\mathcal{O}(N^3)$, because \name{} repeats Gaussian process regression until abnormal signals are all identified.

We solve it by conducting the survey process for each region, thus small-scale training data is used for every site survey. Assuming the number of fingerprints in each region is the same, the computational cost just grows linearly with the survey area rather than in cubic with the survey area. In our experiments, the time needed for signal recovery computation is 14s on the 3th floor and 23s on the 6th floor, which is trivial compared to the site survey time.




\section{Conclusion}
\label{sec:conclusion}

To mitigate the overhead of fingerprint map construction for WiFi-based indoor location systems, the robot are adopted to perform fingerprinting. However, the high time and energy cost make the large deployment of the proposed autonomous fingerprinting systems difficult. \name{} is designed as an energy efficient autonomous fingerprinting system. It conducts the site survey without requiring the robot stop at every reference location, thus saving time and power consumed by de/acceleration. To guarantee the quality of \name{}'s fingerprint relatively small database, two kinds of signal recovery methods are proposed to solve the unreliable signals during the site survey.

We have deployed and evaluated \name{} on two sites of our Department building. The results validate \name's ability in quick finishing the fingerprint map construction and achieving unabated localization accuracy. 

\bibliographystyle{IEEEtran}
\bibliography{bibliography}

\end{document}